\documentclass[
pra,
aps, reprint, longbibliography,superscriptaddress,amsmath,amsfonts
]{revtex4-2}

\usepackage{soul}
\usepackage{graphicx}
\usepackage{dcolumn}
\usepackage{bm}
\usepackage{times}

\usepackage{float}
\makeatletter
\let\newfloat\newfloat@ltx
\makeatother
\usepackage{algorithm}
\usepackage{algpseudocode}
\usepackage{enumitem}
\usepackage{braket}
\usepackage{hyperref}
\usepackage{xcolor}
\usepackage[caption=false]{subfig}

\usepackage{comment}

\begin{document}


\title{Quantum Imaginary Time Evolution on an Infinite 1D Chain}

\author{Hao-Ti Hung}
 
\author{Tung Tsao}%
\author{Ying-Jer Kao}%
\email{yjkao@phys.ntu.edu.tw}
\affiliation{Center for Theoretical Physics, National Taiwan University, Taipei, 10607, Taiwan}
\affiliation{%
 Department of Physics, National Taiwan University, Taipei, 10607, Taiwan
}%





\begin{abstract}
We introduce a quantum-circuit algorithm for performing imaginary-time evolution on infinite one-dimensional lattice systems. The method uses a parameterized quantum circuit to represent a uniform matrix product state  ansatz. We derive the ITE algorithm using the time-dependent variational principle and employ the quantum Lanczos algorithm to improve the ground-state energy estimate. As a benchmark, we simulate the transverse-field Ising model using both classical simulators and IBM Quantum devices. Our analysis includes a statistical study of the distributions of the cost function and energy density obtained from quantum measurements, illustrating the effects of finite-sampling noise on convergence.
\end{abstract}

\maketitle

\section{\label{sec:intro}Introduction}

The advent of noisy intermediate-scale quantum (NISQ) computers~\cite{Preskill2018quantumcomputingin} has created new opportunities to study quantum many-body systems. 
These devices have spurred the development of various quantum algorithms, among which quantum-inspired tensor-network algorithms play a crucial role. 
%
Matrix product states (MPSs)~\cite{Fannes1992,Stellan1995} have proven highly successful for simulating one-dimensional quantum systems~\cite{White1992PRL,White1992PRB,McCulloch_2007,SCHOLLWOCK201196} and can be directly encoded as quantum circuits. Consequently, several classical MPS time-evolution algorithms, such as time-evolving block decimation (TEBD)~\cite{Vidal_2003,Vidal_2004,Vidal_2007} and the time-dependent variational principle (TDVP)~\cite{Haegeman_2011,Haegeman_2016}, have been translated into quantum circuits to simulate quantum dynamics on NISQ devices~\cite{Cramer2010,Ran_2020,Barratt2021,Smith_2022,Astrakhantsev_2023,jamet_2023,maccormack2021}. 

Algorithms for real-time evolution can exploit the unitary nature of quantum gates to simulate quantum dynamics~\cite{Smith2019, Barratt2021,Lin_2021,Mueller_2023}.
However, extending these methods to imaginary-time evolution (ITE), a common approach for finding the ground state of a quantum system, is more challenging. 
The difficulty lies in the fact that the ITE operator is non-unitary, making it incompatible with the inherently unitary nature of quantum gates. 
Three primary approaches have been proposed to address this problem: quantum ITE (QITE), variational ITE (VITE), and probabilistic ITE (PITE)~\cite{Kosugi_2022,Leadbeater_2024}. 
QITE approximates the non-unitary evolution through the use of unitary gates~\cite{Motta2020,Yeter-Aydeniz_2020,Gomes2020,Nishi2021,Sun_2021,Yeter-Aydeniz_2021,Yeter-Aydeniz_2022,Cao2022,Tsuchimochi2023,kumar2024,kolotouros_2024}. 
However, the ground-state energy obtained inevitably contains errors arising from the unitary approximation of the non-unitary operator. 
PITE eliminates this approximation error by embedding the ITE operator in a unitary operator, at the expense of introducing an additional ancilla qubit for each non-unitary operator~\cite{Williams_2004,Hu2020,Lin_2021,Liu2021,Kosugi_2022,nishi_2022,Schlimgen_2022,Turro_2022,Kosugi2023,ManginBrinet2024efficientsolutionof,Xie_2024,Nishi_2024,Leadbeater_2024}. 
Although the ancilla qubits can be reused by measuring and resetting them, so that the total number of qubits does not grow linearly with the evolution time but only depends on the system size, the number of qubits that must be measured does increase linearly with the number of time steps. 
Consequently, the total number of shots required to achieve a given accuracy also grows. 
VITE, on the other hand, employs a variational approach in which the quantum state is represented by a parameterized ansatz, and the parameters are iteratively optimized to simulate ITE~\cite{Jones_2019,McArdle2019,Yuan2019theoryofvariational,Amaro2022,Benedetti_2021,Zhang_2024}. 
The optimization depends on the definition of the cost function. 
If the cost value is obtained directly from the statevector in simulation, it can be computed exactly. 
However, on real quantum devices, the cost must be estimated through repeated measurements, where expectation values are inferred from outcome probabilities.
This measurement process introduces statistical fluctuations, which significantly influence the convergence of the variational optimization.

%

Each of these methods provides a practical route to implementing ITE on quantum circuits, but applications have so far been primarily restricted to finite-size systems. 
In this work, we demonstrate how to perform ITE with quantum circuits for an infinite 1D quantum system. 
Our approach is based on TDVP, with the variational parameters of the quantum circuit updated at each time step. 
As in VITE, the number of quantum gates therefore does not grow with the evolution time. 
To implement the non-unitary ITE operator, we explore two strategies: (i) approximating it with a unitary gate, as in QITE, and (ii) introducing an additional ancilla qubit to construct an exact unitary embedding, as in PITE. 
For convenience, we refer to our quantum-circuit ITE method for infinite 1D systems as QITE, although it should be distinguished from the QITE scheme discussed in the previous paragraph.
We present results from both statevector and measurement-based simulations. 
In the latter case, since the cost value is estimated via repeated measurements, it represents a statistical quantity that follows a distribution rather than a fixed value. 
We therefore analyze how the distribution of cost values affects the distribution of the energy density during the evolution. 
Finally, we implement our algorithm on IBM Quantum devices and employ the quantum Lanczos (QLanczos) algorithm~\cite{Motta2020,Yeter-Aydeniz_2020,Yeter-Aydeniz_2021,Yeter-Aydeniz_2022,Tsuchimochi2023} to improve the ground-state energy estimate, showing that the energy approaches the ground-state value during ITE.


The remainder of this paper is organized as follows. In Sec.~\ref{sec:tdvp_cirq}, we review the quantum-circuit TDVP framework of Ref.~\cite{Barratt2021} and its application to real-time evolution in infinite 1D quantum systems. In Sec.~\ref{sec:qite}, we present two approaches, analogous to QITE and PITE, for implementing ITE in infinite systems and describe how QLanczos is used to refine the results. Section~\ref{sec:res} presents statevector and measurement-based simulations, including implementations on real IBM Quantum devices. In Sec.~\ref{sec:statistics}, we discuss the statistics of the cost values and energy densities obtained from measurement-based estimates. Finally, Sec.~\ref{sec:conclusion} summarizes our findings.

\section{\label{sec:tdvp_cirq}TDVP and quantum circuits}
In this section, we review the main ideas of Ref.~\cite{Barratt2021}, focusing on how TDVP can be used to perform real-time evolution with a quantum circuit. We first introduce TDVP for uniform matrix product states (uMPSs)~\cite{Haegeman_2011,Laurens_2019} and then reformulate it in terms of quantum circuits for implementation on quantum devices.

\subsection{\label{sec:tdvp_umps}uMPS-based TDVP}
Assuming translational invariance, a quantum state $|\psi \rangle$ can be described by the same tensor at every lattice site. In general, a uMPS can be expressed as
\begin{equation}
    \label{eq:uMPS}
    |\psi\rangle=\sum_{\{i\},\{s\}}{\prod_{n}{\cdots A_{i_{n-1},i_n}^{s_{n-1}}A_{i_{n},i_{n+1}}^{s_n}\cdots|\cdots s_{n-1}s_{n}\cdots\rangle}},
\end{equation}
where all rank-3 tensors $A$ are identical.
The corresponding tensor diagram is
\begin{equation}
    \includegraphics[scale=0.75]{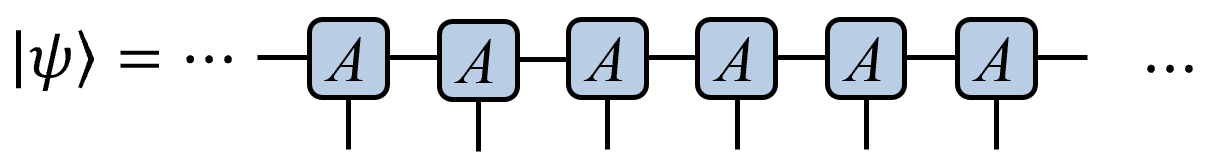}.
	\centering
	\label{fig:uMPS}
\end{equation}
The uMPS spans a manifold within the entire Hilbert space. 
We denote the uMPS by $|\psi(X(t))\rangle$, where $X(t)$ is the set of parameters defining the state at time $t$. After a short real-time interval $dt$ under the Hamiltonian $H$, the state is $e^{-iHdt}|\psi(X(t))\rangle$. The goal is to identify a uMPS $|\psi(X(t+dt))\rangle$ that closely approximates this evolved state by solving
\begin{equation}
\label{eq:opt}
    X(t+dt)=\mathop{\arg\max}_{X(t+dt)}{\langle\psi(X(t+dt))|e^{-iHdt}|\psi(X(t))\rangle}.
\end{equation}
This optimization can be regarded as a projection onto the uMPS manifold 
and embodies the central idea of TDVP~\footnote{Unlike the projection onto the tangent space of the uMPS described in Ref.~\cite{Haegeman_2011}, this approach involves a direct projection onto the manifold.}.

\begin{figure}[htbp]
    \includegraphics[scale=0.8]{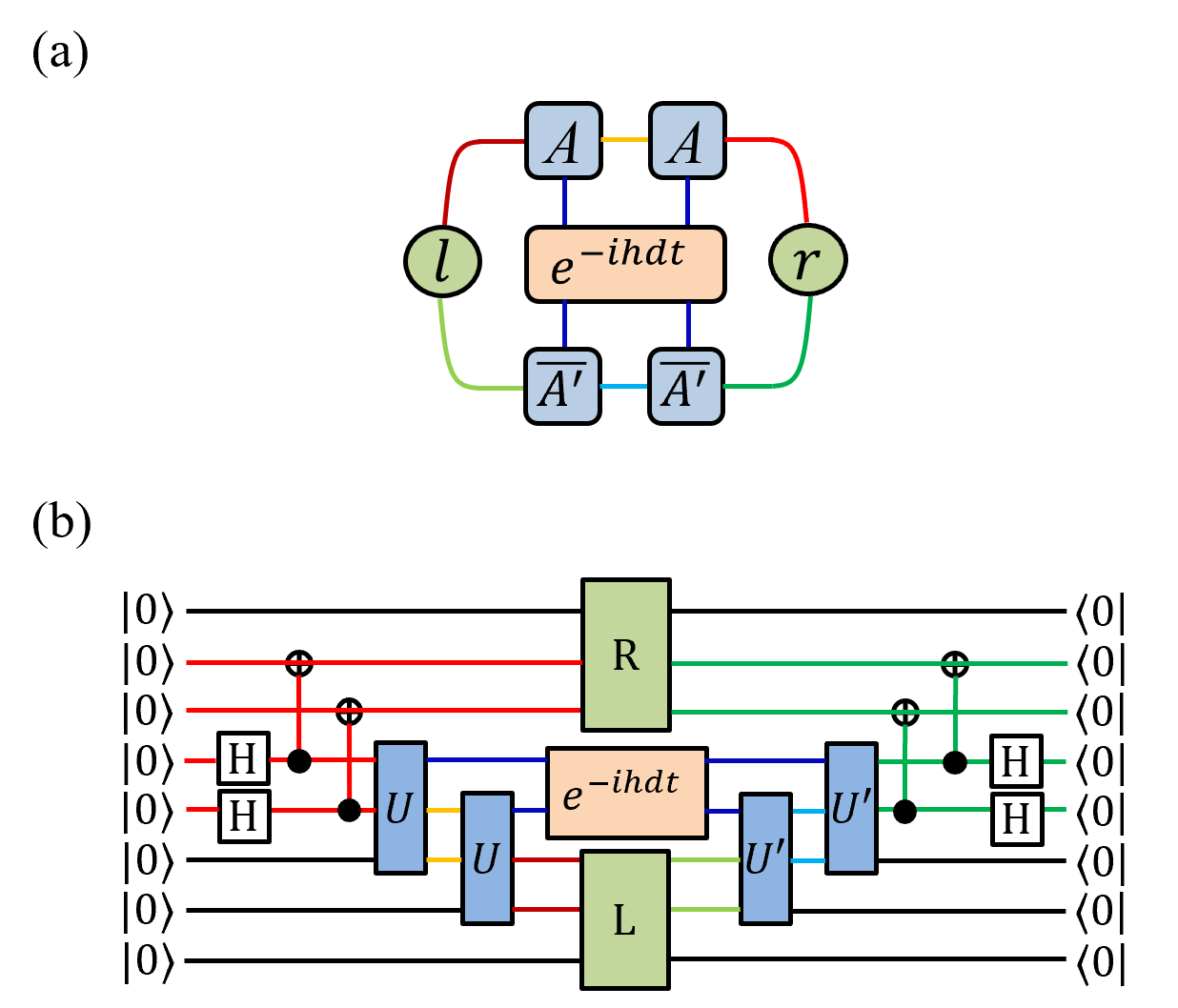}
	\centering
	\caption{    Transformation of the cost function in Eq.~\eqref{eq:opt} from a TN diagram to a quantum-circuit representation. Because the bra state $\langle000\cdots|$ cannot be directly implemented on a quantum computer, the overlap is estimated by measuring the output state.
    }
	\label{fig:cost_uMPS_to_qc}
\end{figure}

If the Hamiltonian contains only nearest-neighbor terms, $H=\sum_{n}{h^{[n,n+1]}}$, the time-evolution operator $e^{-iHt}$ can be approximated using a Suzuki--Trotter decomposition:
\begin{equation}
\label{eq:trotter}
     \left( e^{-i \sum_{n\in\text{odd}} h[n,n+1] \Delta t} \cdot e^{-i \sum_{n\in\text{even}} h[n,n+1] \Delta t} \right)^{m},
\end{equation}
where $\Delta t=t/m$. In tensor-network (TN) language, this expression corresponds to alternating layers of operators acting on odd and even bonds. Translational invariance allows us to consider a single layer. Consequently, Eq.~\eqref{eq:opt} can be represented by the following infinite-chain tensor diagram:
\begin{equation}\label{eq:cost_diag}
    \includegraphics[scale=0.67]{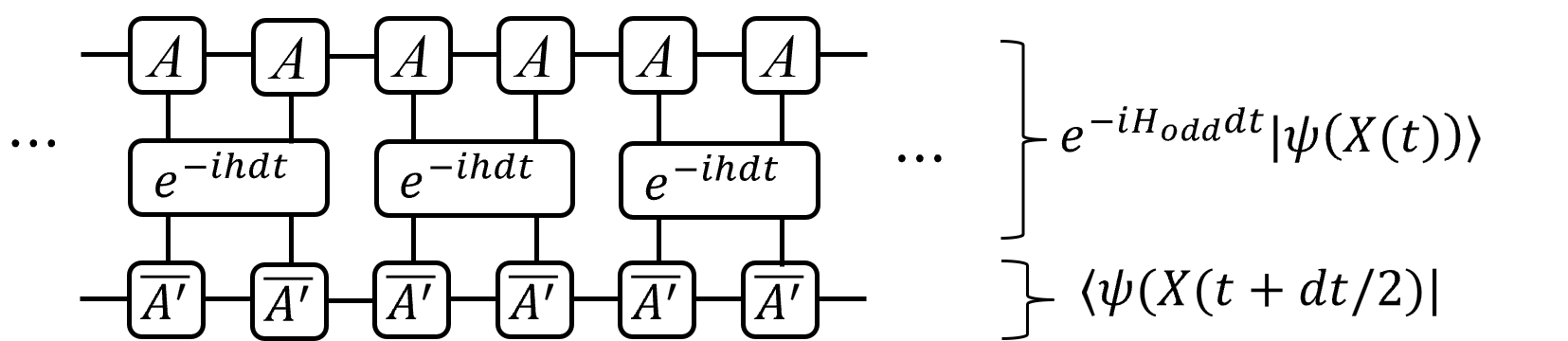}.
\end{equation}
Each layer evolves the system by only $dt/2$, because a full time step $dt$ requires two applications of the layer. 
The infinite chain in Eq.~\eqref{eq:cost_diag} can be simplified by introducing environment tensors. In Fig.~\ref{fig:cost_uMPS_to_qc}(a), the tensors $l$ and $r$ represent the left and right environments, which are obtained by solving fixed-point equations (see Appendix~\ref{appdx:env_qc} for details). The resulting diagram contains only seven tensors, as illustrated in Fig.~\ref{fig:cost_uMPS_to_qc}(a).

\subsection{\label{sec:mapping}Mapping between uMPS and quantum circuits}
To perform simulations on quantum devices, the cost function in Eq.~\eqref{eq:opt} must be represented in the form of a quantum circuit rather than a tensor diagram (see Fig.~\ref{fig:cost_uMPS_to_qc}(a)). 
This requires a mapping from the uMPS tensors to unitary gates in a quantum circuit. 
\label{sec:uMPStoQC}
\begin{figure}[ht]
    \includegraphics[scale=0.93]{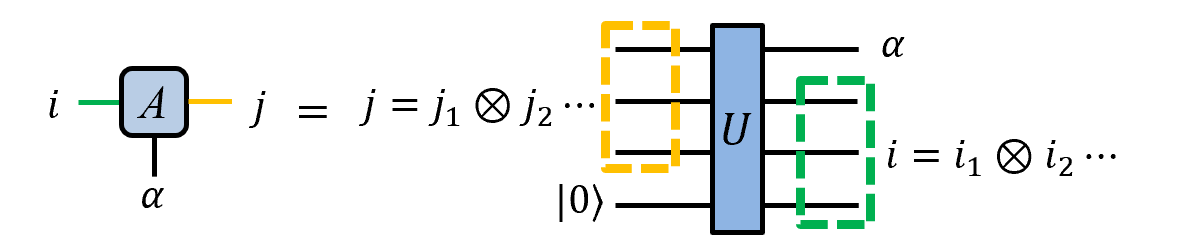}
	\centering
	\caption{The mapping between a one-site unit cell of the uMPS and a unitary gate in the quantum circuit.}
	\label{fig:umps_to_qc}
\end{figure}
As illustrated in Fig.~\ref{fig:umps_to_qc}, a uMPS with virtual bond dimension $D=2^N$ and physical dimension $d=2$ can be mapped to a unitary gate acting on $N+1$ qubits. 
This mapping between uMPS and quantum circuits automatically enforces the canonical form of the tensors~\cite{Barratt2021}. 
The environment tensors $l$ and $r$ are generally not unitary matrices and therefore cannot be directly represented as quantum gates. 
However, by enlarging the Hilbert space, we can construct unitary gates that encode the required environment tensors.
\begin{figure}[ht]
	\includegraphics[scale=1]{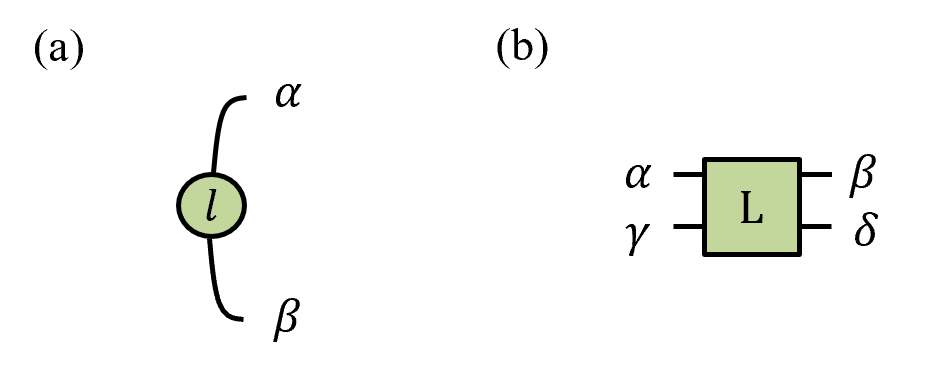}
 \centering
    \caption{
    (a) The left environment of the uMPS.
    (b) The left environment of the quantum circuit.
    }
    \label{fig:l_L}
\end{figure}
Consider the left environment tensor $l_{\alpha,\beta}$ as a $D\times D$ matrix. To encode this non-unitary tensor as a unitary gate, we enlarge it to a $2D\times 2D$ tensor $L_{(\alpha,\gamma),(\beta,\delta)}$, where $\gamma,\delta\in\{0,1\}$. The resulting gate must satisfy two conditions:
\begin{equation}
    \begin{cases}
      \text{$L$ is unitary, namely} & L L^\dagger=I\\
      L_{\alpha,\gamma=0,\beta,\delta=0}=l_{\alpha,\beta}
    \end{cases}    
    \label{eq:L_cond}.
\end{equation}
The first condition ensures that $L$ is unitary, while the second embeds the original environment tensor $l$ in the enlarged space. The right environment gate $R$ must satisfy the analogous condition for $r$. Details of this construction are given in Appendix~\ref{appdx:expand_non_u}.
After the environment gates have been constructed, the cost function in Eq.~\eqref{eq:opt} can be represented entirely as a quantum circuit. The circuit, shown in Fig.~\ref{fig:cost_uMPS_to_qc}(b), includes the unitary gates derived from both the uMPS and its environment tensors and will be used for the quantum-circuit ITE simulations.

\subsection{\label{sec:rte}Real-time evolution of an infinite 1D system}
\begin{figure}[t]
    \includegraphics[scale=1]{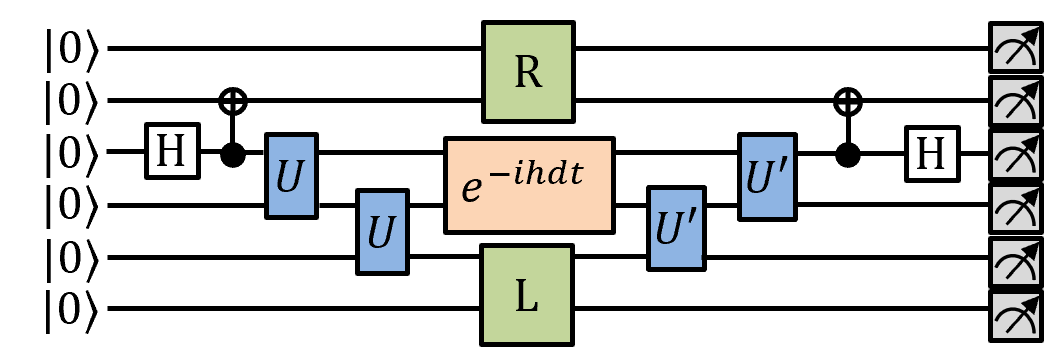}
	\centering
	\caption{Quantum-circuit representation of the cost function in Eq.~\eqref{eq:opt} for bond dimension $D=2$.}
	\label{fig:cost_circ_meas}
\end{figure}

This construction applies to any bond dimension $D=2^N$. Here, we consider $D=2$, for which the cost-function circuit requires only six qubits. The cost function in Fig.~\ref{fig:cost_uMPS_to_qc} is the overlap of the circuit output with $\langle000\cdots|$ and can therefore be estimated from the probability of measuring the all-zero output state, as illustrated in Fig.~\ref{fig:cost_circ_meas}. 

\begin{figure*}
    \includegraphics[scale=0.8]{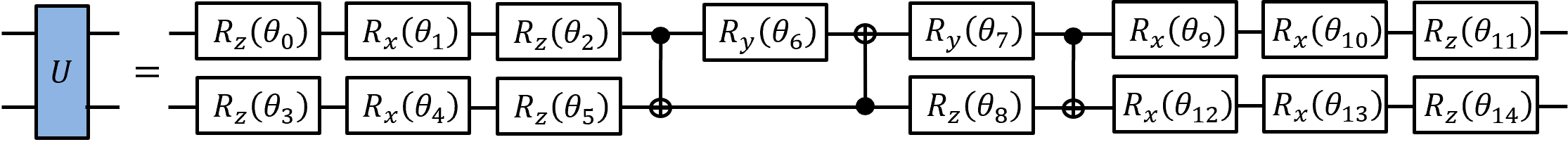}
	\centering
	\caption{Example of a PQC with virtual bond dimension $D=2$. We use the ansatz provided in the GitHub repository associated with Ref.~\cite{Barratt2021}. Here, $R_i$ denotes a rotation about the $i$ axis.}
	\label{fig:ansatz}
\end{figure*}

To perform real-time evolution, we parametrize the unitary gate $U$ in Fig.~\ref{fig:cost_uMPS_to_qc}(b) using the parameterized quantum circuit (PQC) ansatz shown in Fig.~\ref{fig:ansatz}. The gate $U$ represents the current state $|\psi(X(t))\rangle$. Our goal is to find the optimized gate $U'$ corresponding to $|\psi(X(t+dt/2))\rangle$ by maximizing the overlap represented by the circuit in Fig.~\ref{fig:cost_uMPS_to_qc}(b). The cost function is evaluated on a quantum simulator or a real quantum device, whereas the optimization of $U'$ is performed on a classical computer. 
After finding the optimal $U'$ for the current time step, we replace $U$ with $U'$ and repeat the procedure to track the quantum dynamics. Reference~\cite{Barratt2021} demonstrated that this approach can perform statevector-based real-time evolution and accurately reproduce the dynamical quantum phase transition (DQPT)~\cite{Heyl_2013, Heyl_2018} of the transverse-field Ising model.


\begin{figure*}
	\includegraphics[scale=0.7]{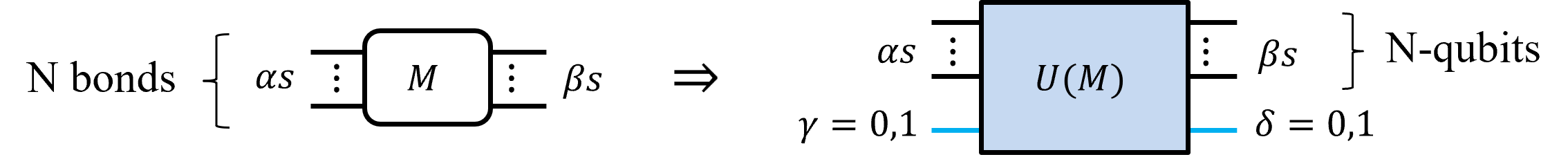}
 \centering
    \caption{A non-unitary operator can be embedded in a unitary gate by introducing an additional qubit, commonly called an ``ancilla qubit.''
    }
    \label{fig:non_u_trans}
\end{figure*}


\section{\label{sec:qite}Quantum imaginary-time evolution for an infinite system}
Although real-time evolution can be implemented naturally using unitary quantum circuits, ITE is more challenging because its evolution operator is non-unitary. Implementation on real quantum devices introduces further difficulties. In this section, we present two strategies for handling the non-unitary ITE operator and describe how QLanczos can improve the resulting energy estimates.
\subsection{\label{sec:non_unitary}Two approaches for handling non-unitary operators}
We consider two approaches to the non-unitarity of ITE. The first approximates the non-unitary gate with a unitary one, as in the QITE algorithm~\cite{Motta2020}. Specifically, we identify a time-dependent Hermitian operator $A(\tau)$ and construct a unitary operator $e^{-iA(\tau)d\tau}$ that satisfies
\begin{equation} \label{eq:A_approx}
e^{-iA(\tau)d\tau}|\psi(\tau)\rangle \approx e^{-Hd\tau}|\psi(\tau)\rangle.
\end{equation}
Because non-unitary operators do not preserve the state norm, we determine $A(\tau)$ by minimizing the difference between the normalized ITE state and its unitary approximation:
\begin{equation} \label{eq:A_eq}
\begin{split}
&A(\tau) = \\&\mathop{\arg\min}_{A(\tau)}{\|\frac{e^{-Hd\tau}|\psi(\tau)\rangle}{\sqrt{\langle\psi(\tau)|e^{-2Hd\tau}|\psi(\tau)\rangle}}-e^{-iA(\tau)d\tau}|\psi(\tau)\rangle\|}.
\end{split}
\end{equation}
Once $A(\tau)$ has been found, the non-unitary ITE operator $e^{-Hd\tau}$ is replaced by $e^{-iA(\tau)d\tau}$, ensuring that every gate in the circuit shown in Fig.~\ref{fig:cost_uMPS_to_qc}(b) is unitary. Algorithm~\ref{alg:QITE} summarizes the procedure. We solve the optimization problem for $A(\tau)$ using tensor-network techniques, as detailed in Appendix~\ref{appdx:approch_non_u}.


\newlist{steps}{enumerate}{1}
\setlist[steps, 1]{label = Step \arabic*:}

\begin{algorithm}
\caption{QITE algorithm}\label{alg:QITE}
\begin{steps}
  \item Given the current gate $U$, determine the environments $L$ and $R$, and construct the cost function in Eq.~\eqref{eq:A_eq} to find the optimal operator $A(\tau)$.
\item Construct the cost function using the quantum circuit shown in Fig.~\ref{fig:cost_uMPS_to_qc}(b), replacing $e^{-iHdt}$ with $e^{-iA(\tau)d\tau}$. Find the optimal $U^\prime$ that maximizes this cost function.
\item Update $U\leftarrow U^\prime$.
\end{steps}
\end{algorithm}

The second method, similar to PITE, introduces an additional ancilla qubit and embeds the non-unitary ITE operator in a larger unitary operator. This enlarges the system from $n$ to $n+1$ qubits. For bond dimension $D=2$, the ITE circuit therefore requires seven qubits, rather than the six required for real-time evolution or for the preceding approximation method. A non-unitary operator $M$ can be implemented through the block encoding~\cite{Martyn_2021, Nishi_2024}
\begin{equation} \label{eq:UofM}
U(M)=\begin{pmatrix}
M & \sqrt{I - M^2} \\
\sqrt{I - M^2} & -M
\end{pmatrix}.
\end{equation}
Here, $\sqrt{I-M^2}$ is defined using the singular value decomposition (SVD) of $M$. One can verify that $U^{\dagger}U=I$ provided that the singular values of $M$ do not exceed 1, a condition that can always be achieved by rescaling $M$ to $M/\alpha$.
When $M$ is the ITE operator, this rescaling does not affect the optimization of the cost function used to obtain the new state $|\psi(X(\tau+d\tau/2))\rangle$. The same method can be used to construct the environment gates $L$ and $R$ by setting $M=l$ and $M=r$, respectively, as described in Sec.~\ref{sec:mapping}. Because the singular values of $l$ and $r$ are at most 1, no rescaling is required in these cases. 


\subsection{\label{sec:qlan}Quantum Lanczos algorithm}
The quantum Lanczos (QLanczos) algorithm, introduced in Ref.~\cite{Motta2020}, addresses limitations of QITE, including errors caused by the unitary approximation in Eq.~\eqref{eq:A_approx}, and can converge more rapidly than QITE. We use QLanczos to refine our simulation results. Unlike the traditional Lanczos method, which generates a Krylov subspace from vectors such as $|\phi\rangle$, $H|\phi\rangle$, and $H^2|\phi\rangle$, QLanczos constructs a subspace directly from the ITE sequence $\ket{\phi_l}=e^{-ld\tau H}\ket{\phi_0}$. A regularization parameter $s$ is used to exclude candidate basis states whose overlap with a previously selected state exceeds the threshold $s$. After selecting the basis, we construct the Hamiltonian matrix $H_{l,l'}=\langle\phi_l|\hat{H}|\phi_{l'}\rangle$ and overlap matrix $S_{l,l'}=\langle\phi_l|\phi_{l'}\rangle$. Solving the generalized eigenvalue problem $\bm{H}x=E\bm{S}x$ then yields an approximation to the ground-state energy from its smallest eigenvalue. Algorithm~\ref{alg:QLan} summarizes the procedure.

\begin{algorithm}
\caption{Quantum Lanczos algorithm}\label{alg:QLan}
\begin{steps}
  \item Given the QITE states at different time steps, choose $\ket{\phi_0}=\ket{\phi_{l=0}}$ as the first basis vector.
\item Find the next state $\ket{\phi_l}$ satisfying $|\langle\phi_l|\phi_0\rangle|<s$, where $0<s<1$ is a regularization parameter, and add it to the selected basis $\{|\Phi_l\rangle\}$.
\item Set $\ket{\phi_0}\leftarrow\ket{\phi_l}$ and repeat Step 2 until the desired number of basis vectors has been selected. 
\item Calculate $S_{l,l'}=\langle\Phi_l|\Phi_{l'}\rangle$ and $H_{l,l'}=\bra{\Phi_l}\hat{H}\ket{\Phi_{l'}}$, and solve $\bm{H}x=E\bm{S}x$. The smallest eigenvalue $E$ approximates the ground-state energy. 
\end{steps}
\end{algorithm}

In our simulations, we save the unitary gates at each time step and convert them to the unit-cell form of a uMPS (see Fig.~\ref{fig:umps_to_qc}). We compute the overlap matrix elements $S_{l,l'}$ from the leading eigenvalues of the corresponding uMPS transfer matrices. To obtain the Hamiltonian matrix elements $H_{l,l'}$, we use standard tensor-network techniques and construct a network analogous to that in Fig.~\ref{fig:cost_uMPS_to_qc}(a), with the real-time evolution operator replaced by the local Hamiltonian $h$. Section~\ref{sec:res} discusses the resulting QLanczos refinement in detail.

\section{\label{sec:res}Simulation results}
We now present simulation results for an infinite 1D transverse-field Ising model (TFIM) at its critical point, with Hamiltonian
\begin{equation} \label{eq:Ising}
    H=\sum_i\left({-\sigma_i^z\sigma_{i+1}^z+\sigma_i^x}\right).
\end{equation}
The exact critical ground-state energy density is $-4/\pi\simeq-1.2732$, providing context for the $D=2$ variational reference energy $-1.2725$ used below.

\begin{figure*}[htbp]
  \centering
  \includegraphics[width=\linewidth]{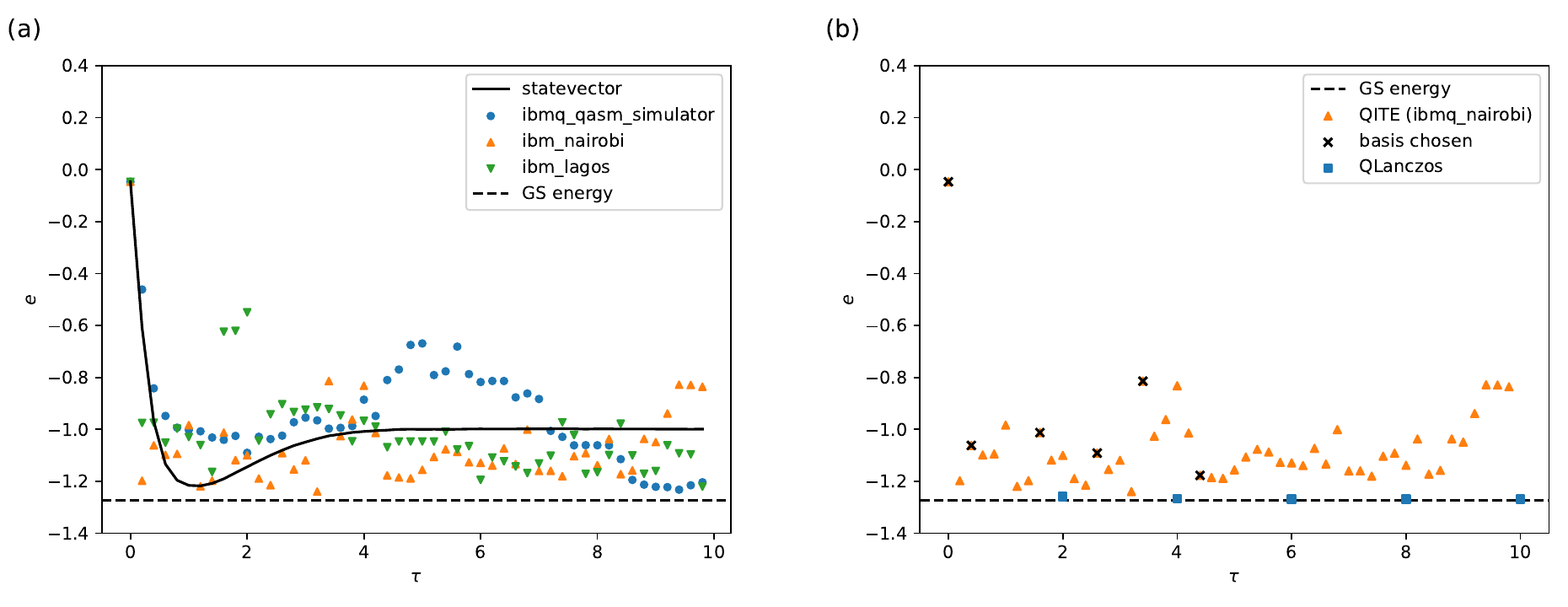}
\caption{(a) ITE results from a simulator and real IBM Quantum devices using the unitary approximation $e^{-iA(\tau)d\tau}$. The solid black line shows the statevector result, the blue circles show results from the \texttt{ibmq\_qasm\_simulator}, and the upward- and downward-pointing triangles show results from two real IBM Quantum devices. The dashed black line indicates the ground-state energy obtained using VUMPS with virtual bond dimension $D=2$. (b) QLanczos results with regularization parameter $s=0.95$. The orange triangles show results from \texttt{ibmq\_nairobi}, the crosses indicate the selected Krylov basis states, and the blue squares show the QLanczos energies.
}
\label{fig:res}
\end{figure*}

We apply the algorithms described in the previous section with bond dimension $D=2$. The first approach approximates the ITE operator by the unitary operator $e^{-id\tau A(\tau)}$, where the Hermitian operator $A(\tau)$ is obtained by solving Eq.~\eqref{eq:A_eq}. The results are presented in Fig.~\ref{fig:res}. The solid black line denotes the statevector simulation. The energy unexpectedly begins to increase near $\tau\approx1$, a phenomenon also reported in Refs.~\cite{Motta2020,Nishi2021,Cao2022}. This behavior may arise from the limited correlation-domain size~\cite{Motta2020}, which is restricted to two sites in our infinite-system simulations. Unlike in finite-size systems, this domain cannot be enlarged within our present setup.

To reduce the noise and estimation errors associated with the unitary approximation, we apply the QLanczos algorithm described in Sec.~\ref{sec:qlan}. In Fig.~\ref{fig:res}(a), the colored markers show results from quantum simulators and real IBM Quantum devices. We estimate the cost function in Eq.~\eqref{eq:opt}, with $H$ replaced by $A(\tau)$, from measurement probabilities using 8192 shots ($2^{13}$). The cost function is implemented with the six-qubit circuit shown in Fig.~\ref{fig:cost_circ_meas}, with $h$ replaced by $A(\tau)$. Whereas the statevector simulation yields the cost exactly, the measurement-based approach introduces statistical noise. After QLanczos refinement, the results in Fig.~\ref{fig:res}(b) closely match the energy obtained using the classical VUMPS algorithm~\cite{PhysRevB.97.045145} with bond dimension $D=2$. The black crosses indicate the basis states selected for QLanczos. The final state is selected at $\tau=4.2$; no additional basis states are selected thereafter, and the QLanczos energy remains unchanged.


\begin{figure*}[htbp]
  \centering
  \includegraphics[width=\linewidth]{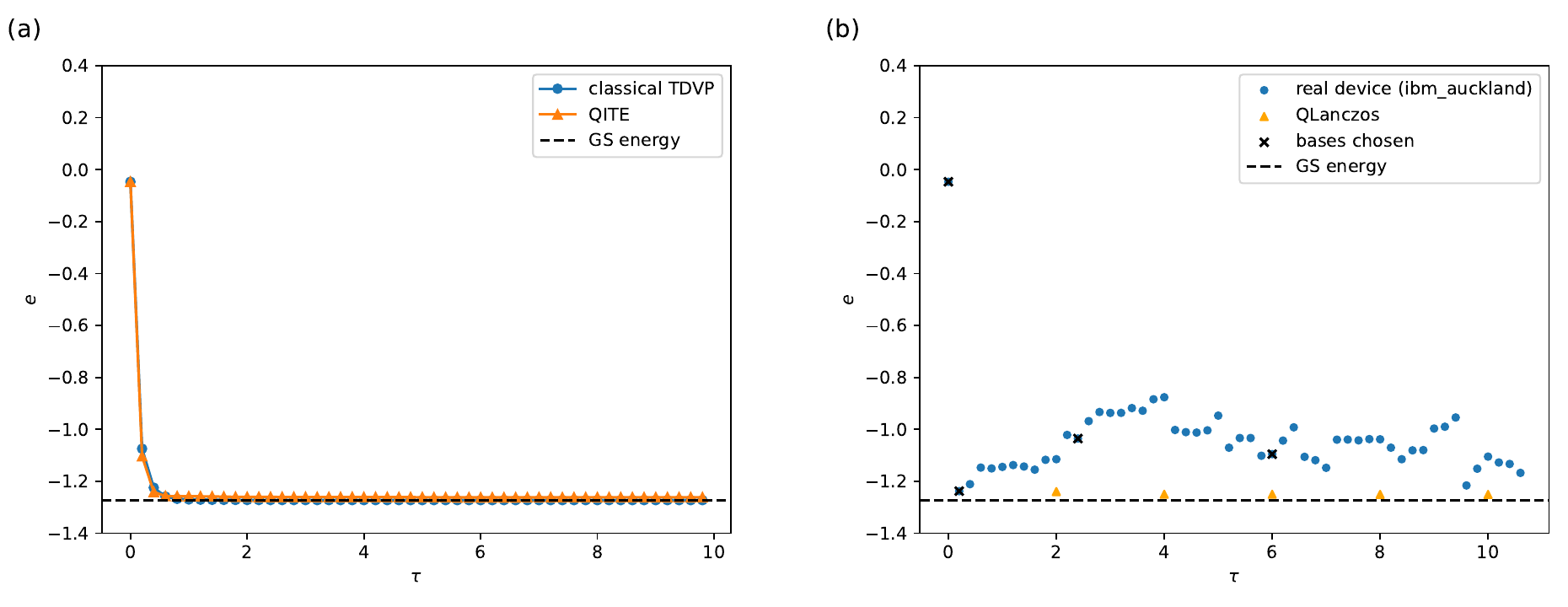}
\caption{(a) Results from the classical TN TDVP algorithm with bond dimension $D=2$ (blue dots) and from the ancilla-assisted unitary embedding of the ITE operator described in Sec.~\ref{sec:non_unitary} (orange markers, labeled QITE in the legend). (b) Results from the real quantum device \texttt{ibmq\_auckland} after QLanczos refinement.}
\label{fig:res2}
\end{figure*}


We also implement the ancilla-assisted unitary embedding of the non-unitary ITE operator. This method requires a seven-qubit circuit instead of the six-qubit circuit used above. The orange triangles in Fig.~\ref{fig:res2}(a) show the statevector results. For comparison, the blue points show classical TN TDVP results with bond dimension $D=2$, computed using the Cytnx library~\cite{Cytnx_2025}; the two sets of results closely agree. We also perform simulations on a real IBM Quantum device. As expected, the results are noisy because the cost function must be estimated from measurements. Nevertheless, the QLanczos refinement shown in Fig.~\ref{fig:res2}(b) substantially reduces the error and brings the energy closer to the ground-state value.



Both methods begin from the ansatz in Fig.~\ref{fig:ansatz} with all parameters $\{\theta_i\}$ set to zero. Although the cost functions are evaluated on real quantum devices in Figs.~\ref{fig:res} and \ref{fig:res2}(b), all optimizations are performed on a classical computer. We use a time step $d\tau=0.2$, evolving to $\tau=10$ (50 time steps per trajectory), and perform the optimization with \texttt{scipy.optimize.minimize} using the \texttt{COBYLA} method and a tolerance of approximately $10^{-3}$. In the first method, which approximates the ITE operator by a unitary operator, the energy begins to increase at long evolution times. The ancilla-assisted method does not exhibit this behavior and converges toward the reference energy. This contrast suggests that the unexpected increase is caused by the unitary-approximation error. Appendix~\ref{appdx:approch_non_u} provides further discussion.

\section{\label{sec:statistics}Statistical effects of quantum measurements}

\begin{figure*}[htbp]
  \centering
  \includegraphics[width=\linewidth]{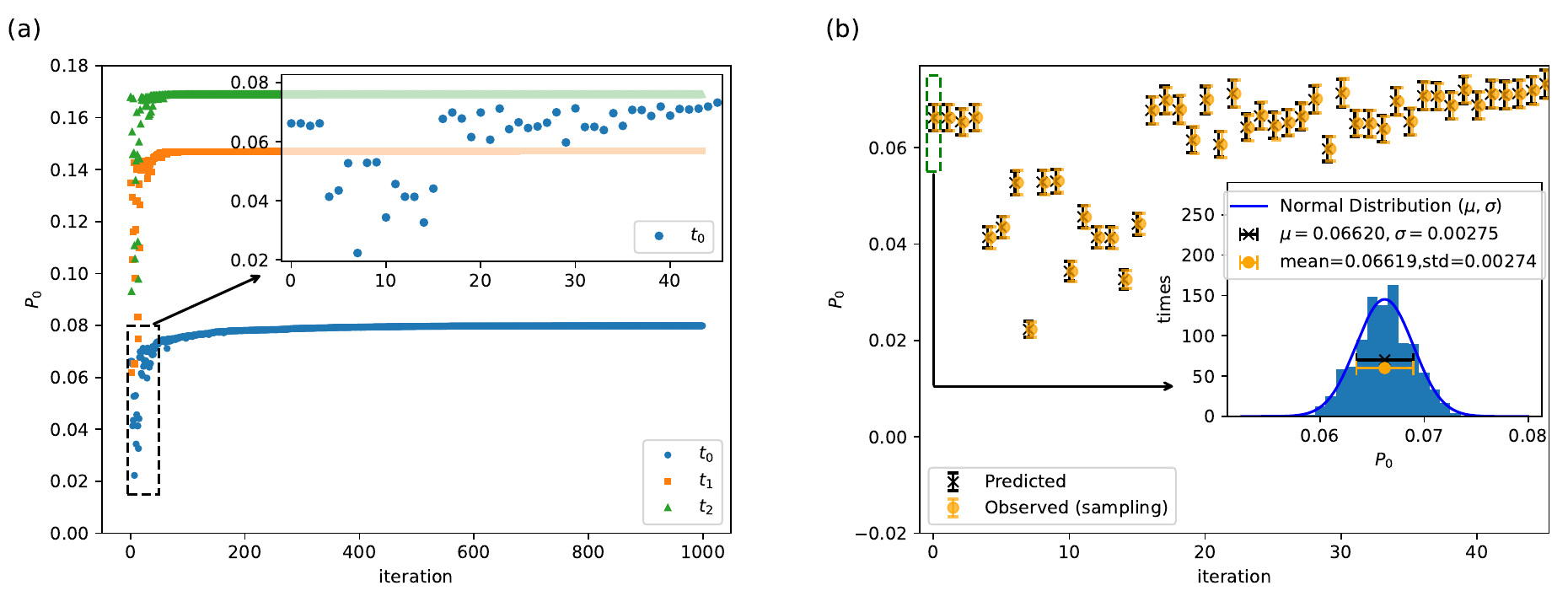}
\caption{
(a) Cost values during the maximization iterations, obtained directly from the statevector for the first three QITE time steps in Fig.~\ref{fig:res2}(a). 
The inset enlarges the region enclosed by the dashed rectangle. 
(b) Comparison of the measured means and standard deviations with predictions based on the statevector probability $P_0$ and Eq.~\eqref{eq:std} for the first QITE time step and the first few maximization iterations, using $N_{\mathrm{shots}}=8192$ and $1000$ independent repetitions. 
The inset shows the histogram obtained from $1000$ repetitions for the first iteration. 
In the inset, the orange error bar with a center dot represents the result from measurement statistics, while the black error bar with a cross marker denotes the prediction from $P_0$ and Eq.~\eqref{eq:std}.
}

\label{fig:statevec_statistics}
\end{figure*}
The quantum-device results in Fig.~\ref{fig:res2}(b) do not converge to the ground-state energy during ITE without further QLanczos refinement. Although current quantum devices are affected by substantial hardware noise, an important limitation here is the statistical error introduced when the cost function is estimated from circuit measurements, as in Fig.~\ref{fig:cost_circ_meas}. The ITE circuit also requires an additional ancilla qubit for this evaluation. In this section, we examine how measurement-based cost estimates affect energy convergence during ITE.

As shown in Fig.~\ref{fig:cost_circ_meas}, the cost is the overlap of the circuit output with the all-zero state. On a real quantum device, this quantity is estimated from repeated measurements of the output state, yielding the probability $P_0$ of obtaining the all-zero bit string. 
This procedure corresponds to a Bernoulli process, where a ``success'' is defined as a measurement outcome in the all-zero state 
and a ``failure'' otherwise. 
In such a process, the standard deviation of the estimator for $P_0$ is given by
\begin{equation}\label{eq:std}
    \sigma = \sqrt{\frac{P_0 (1 - P_0)}{N_{\mathrm{shots}}}},
\end{equation}
where $N_{\mathrm{shots}}$ denotes the number of measurement shots. 
Equation~\eqref{eq:std} describes only the statistical finite-sampling error of the estimator. On real hardware, readout errors and decoherence also introduce systematic bias in $P_0$, so Eq.~\eqref{eq:std} should be regarded as a lower bound on the device-level uncertainty. 
Moreover, since $\sigma$ depends on $P_0$ itself, the noise level varies along the optimization trajectory as the circuit parameters change.
Figure~\ref{fig:statevec_statistics} illustrates this effect. Panel (a) shows the statevector cost during the maximization iterations for the first three QITE time steps in Fig.~\ref{fig:res2}(a). For the first QITE time step and the first few maximization iterations, we estimate the cost on a simulator using $N_{\mathrm{shots}}=8192$ and repeat the measurement $1000$ times to determine the mean and standard deviation. Figure~\ref{fig:statevec_statistics}(b) shows that the measured means and standard deviations (orange error bars) closely agree with the statevector probabilities $P_0$ and the standard deviations predicted by Eq.~\eqref{eq:std} (blue error bars). The inset shows the distribution for the first iteration; it is well approximated by a normal distribution with mean $P_0$ and the standard deviation predicted by Eq.~\eqref{eq:std}.

\begin{figure}[htbp]
  \centering
  \includegraphics[width=\linewidth]{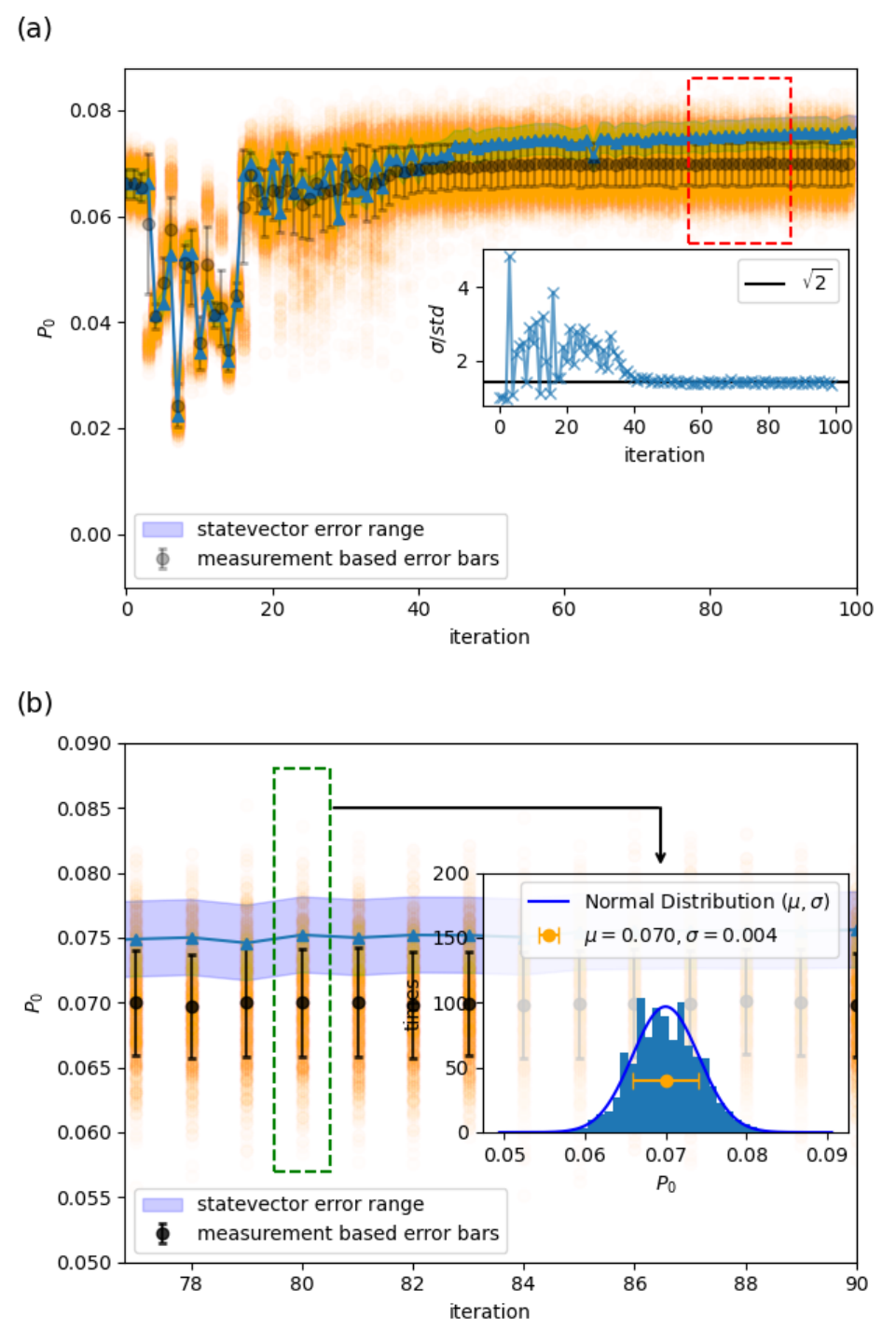}
\caption{(a) Cost values during iterations in which the circuit parameters are updated using measurement-based estimates (orange points with black error bars), obtained from $1000$ repetitions with $N_{\mathrm{shots}}=8192$. 
For comparison, the statevector results are shown as a blue curve with error bars estimated from Eq.~\eqref{eq:std}, also using $N_{\mathrm{shots}} = 8192$. 
The inset shows the ratio of the standard deviation from the measurement-based update to that from the statevector case. This ratio converges toward $\sqrt{2}$ in the saturation regime. 
(b) Enlargement of the red dashed rectangle in panel~(a). 
The inset in (b) shows the distribution of the cost value at the 80th iteration, obtained from $1000$ repetitions. 
The blue line indicates a normal distribution with mean $\mu$ and standard deviation $\sigma$, where $\mu$ is the sample mean of the distribution and $\sigma$ is the standard deviation estimated from the $1000$ samples.
}
\label{fig:meas_update}
\end{figure}

The preceding analysis holds the PQC parameters fixed and samples the cost associated with the statevector probability. We now consider the case in which the circuit parameters themselves are updated using measurement-based cost estimates. The orange points in Fig.~\ref{fig:meas_update}(a) show the costs obtained from $1000$ independent update runs on a simulator with $N_{\mathrm{shots}}=8192$. For comparison, the blue curve shows the statevector results with error bars from Eq.~\eqref{eq:std}; the black error bars indicate the uncertainty of the measurement-based updates. The cost fluctuates strongly during the initial iterations but eventually stabilizes and saturates. 
Nevertheless, the standard deviation in the measurement-based update remains larger than that in the statevector case. 
Interestingly, the ratio between the standard deviation from the measurement-based update and that from the statevector case approaches $\sqrt{2}$ in the saturation regime. 
This ratio indicates that, in the stationary regime, the parameter fluctuations induced by the noisy updates contribute a variance to the measured cost comparable to the direct shot noise of a single estimate, so that the two contributions add in quadrature, $\sigma_{\mathrm{upd}}^2 \approx 2\sigma_{\mathrm{sv}}^2$. 
A quantitative model based on a linearized stochastic description of the update dynamics is given in Appendix~\ref{appdx:sqrt2}.

\begin{figure}[htbp]
  \centering
  \includegraphics[width=\linewidth]{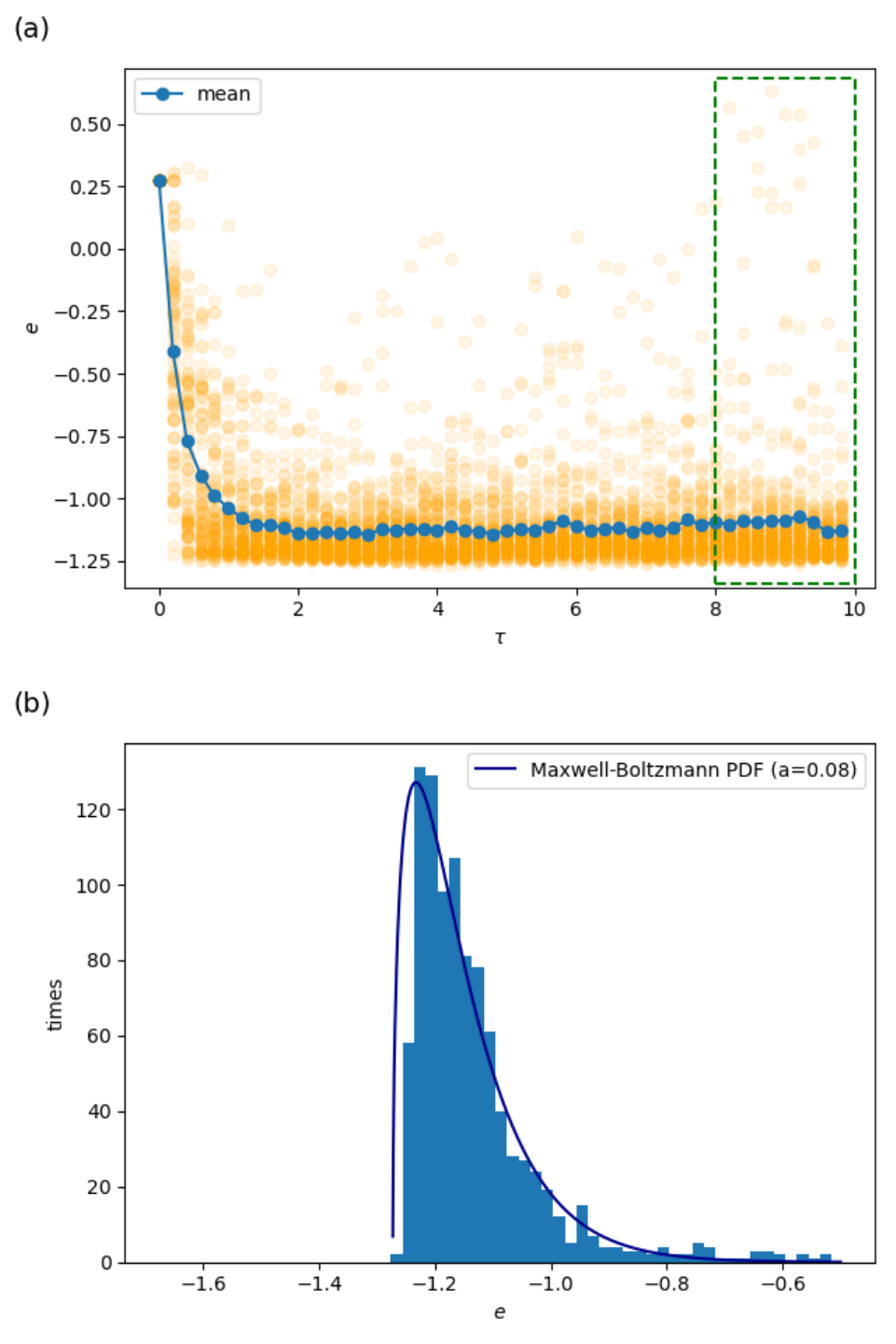}
\caption{
(a) Energy-evolution curves from 100 independent QITE simulations on a noiseless simulator, with circuit parameters updated using measurement-based estimates with $N_{\mathrm{shots}}=8192$.  
(b) Energy distribution obtained from the last 10 time steps of each simulation (1000 data points in total).
}
\label{fig:e_evo_statistics}
\end{figure}
Recall that the real-quantum-device simulation in Fig.~\ref{fig:res2}(b) does not converge to the ground-state energy during the evolution. To investigate this behavior, we repeat the full QITE procedure 100 times on a classical simulator, updating the circuit with measurement-based estimates and using the same initial state for every run. 
This yields 100 energy-evolution curves, as shown in Fig.~\ref{fig:e_evo_statistics}(a). 

In these simulations, the energy approaches the ground-state value and remains bounded from below by it, as required for variational energies. 
Notably, the lowest energy attained across all $100$ runs lies approximately $0.013$ above the $D=2$ variational minimum, indicating that every trajectory stalls before reaching the optimum, at the point where the optimizer tolerance becomes comparable to the finite-sampling noise of the cost estimates. 
We analyze the last 10 time steps of each trajectory, corresponding to the late-time window of the evolution, for a total of 1000 data points. 
The resulting energy distribution, shown in Fig.~\ref{fig:e_evo_statistics}(b), is empirically well approximated by a Maxwell--Boltzmann-like form (a Gamma distribution with shape $3/2$):
\begin{equation}\label{eq:maxwell_boltzmann}
    f(E) = 2\sqrt{\frac{E-E_0}{\pi}}\left(\frac{1}{a}\right)^{3/2} \exp\left(-\frac{E-E_0}{a}\right),
\end{equation}
where $a=0.08$; the parameter $E_0$ is identified with the stall threshold discussed above rather than the ground-state energy.

To clarify the origin of this distribution, we analyze the ensemble of trajectories in more detail in Appendix~\ref{appdx:energy_stat}; here we summarize the main findings. 
The ensemble is not statistically homogeneous: 16 of the 100 runs fail to converge and account for the secondary structure near $e \approx -0.94$ as well as the entire high-energy tail, while among the converged runs the run-to-run spread of stalled plateau energies is comparable to the fluctuations within a single run, so that a substantial fraction of the observed width reflects the distribution of distinct stall points rather than stationary noise about a common minimum. 
Moreover, the fluctuations within an individual run are approximately Gaussian, consistent with the energy varying linearly in the Gaussian parameter noise at a non-stationary stall point, where the local gradient has not vanished. 
The Maxwell--Boltzmann-like form of Eq.~\eqref{eq:maxwell_boltzmann} should therefore be understood as an empirical description of the aggregate envelope---the convolution of a right-skewed distribution of stall energies with approximately Gaussian within-run fluctuations, together with the unconverged subpopulation---rather than as evidence for a single underlying $\chi^2$ law; consistent with this interpretation, the fitted shape parameter of a generalized (Gamma) form is not robust, varying between roughly $1$ and $3$ depending on how the floor $E_0$ and the unconverged tail are treated (see Appendix~\ref{appdx:energy_stat}).

\section{\label{sec:conclusion}Conclusion}

We have introduced quantum-circuit algorithms for simulating ITE in infinite 1D systems. A variational quantum-circuit ansatz represents the projected uMPS state. Because the ITE operator is non-unitary, we consider two implementations. The first approximates it with a unitary operator without enlarging the Hilbert space, but the resulting approximation error causes an unexpected increase in energy at long evolution times. The second introduces an ancilla qubit and embeds the ITE operator in a larger unitary operator. This method doubles the operator dimension and agrees closely with the classical TN TDVP results.

Although the ancilla-assisted method performs well in statevector simulations, estimating the cost function through finite sampling introduces statistical errors on quantum devices. The measured cost values are approximately normally distributed, whereas the aggregate energy distribution forms a right-skewed, Maxwell--Boltzmann-like envelope. Our trajectory-resolved analysis attributes this envelope to run-to-run variations among stalled optimization points above the variational minimum, combined with approximately Gaussian fluctuations within each run. QLanczos post-processing mitigates these errors and produces a more accurate ground-state energy estimate.

For bond dimension $D=2$, our uMPS implementation requires only seven qubits. Although we focus on this case, recent studies have explored mappings between MPSs, including uMPSs, and quantum gates at larger bond dimensions~\cite{Ran_2020,Lin_2021,Astrakhantsev_2023}. Our method can therefore, in principle, be generalized to larger bond dimensions to simulate more complex and strongly entangled systems. Such extensions are a promising direction for future work.

\begin{acknowledgments}
We are grateful to the IBM Q Hub at NTU for providing simulation resources on the IBM Q System. We  acknowledge the use of Claude (Anthropic) and ChatGPT (OpenAI) for assistance with editing, and literature search. This work is partially supported by NSTC through grants No. 113-2112-M-002-033-MY3, 115-2124-M-001-015- and also by grant NSF PHY-2309135 to the Kavli Institute for Theoretical Physics (KITP).
\end{acknowledgments}

\section*{Data availability}
The energy-trajectory data underlying Figs.~\ref{fig:e_evo_statistics} and \ref{appdx_fig:e_stat}, together with the analysis scripts used in Appendix~\ref{appdx:energy_stat}, are available from the corresponding author upon reasonable request.

\bibliography{ref}
\newpage
\appendix
\section{Unitary embedding and environment gates}
\label{appdx:expand_non_u}
\label{appdx:env_qc}

Any non-unitary $n\times n$ operator $M$ can be embedded in a larger $2n\times2n$ unitary matrix of the form
\begin{equation}
U=
\begin{pmatrix}
M & A \\
B & C 
\end{pmatrix}.
\end{equation}
We perform the singular value decomposition (SVD)
\begin{equation}
M=WSV^{\dagger},
\end{equation}
and construct the unitary operator
\begin{equation}
U=
\begin{pmatrix}
M & W\sqrt{I-S^2}V^{\dagger} \\
W\sqrt{I-S^2}V^{\dagger} & -M 
\end{pmatrix}.
\end{equation}
We can directly verify that $U$ is unitary:
\begin{widetext}
\begin{equation}
\begin{split} 
&UU^\dagger \\& =
\begin{pmatrix}
WSV^{\dagger} & W\sqrt{I-S^2}V^{\dagger} \\
W\sqrt{I-S^2}V^{\dagger} & -WSV^{\dagger} 
\end{pmatrix}
\begin{pmatrix}
V SW^\dagger & V\sqrt{I-S^2}W^\dagger \\
V\sqrt{I-S^2}W^\dagger & -V SW^\dagger 
\end{pmatrix} 
\\
&=
\begin{pmatrix}
WS^2W^\dagger + W(I-S^2)W^\dagger & 0 \\
0 & WS^2W^\dagger + W(I-S^2)W^\dagger  
\end{pmatrix} \\
&=I\\
\end{split}.
\end{equation}
\end{widetext}
Thus, a non-unitary operator $M$ acting on $N$ qubits (see Fig.~\ref{fig:non_u_trans}) can be embedded in a unitary operator $U(M)$ acting on $N+1$ qubits. We use this construction both to obtain the environment gates $L$ and $R$ and to implement the imaginary-time evolution operator.

\begin{figure}[t]
    \includegraphics[width=\columnwidth]{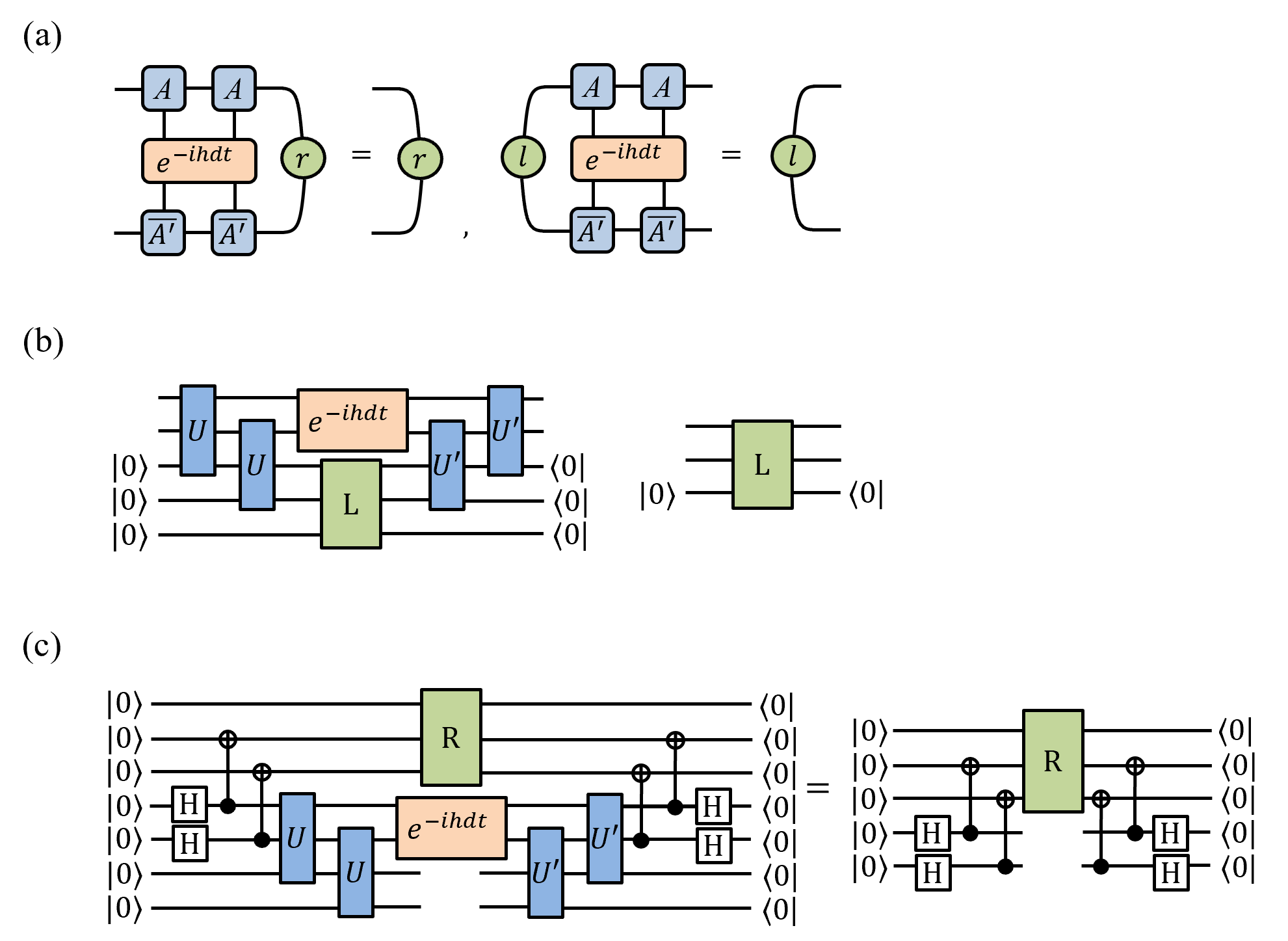}
	\centering
	\caption{
        (a) Fixed-point equations for the uMPS with the evolution operator.
        (b) and (c) Corresponding fixed-point equations for the quantum circuits.
    }
	\label{appdx_fig:unitary_env}
\end{figure}

\begin{figure*}[t]
    \centering
    \includegraphics[width=\textwidth]{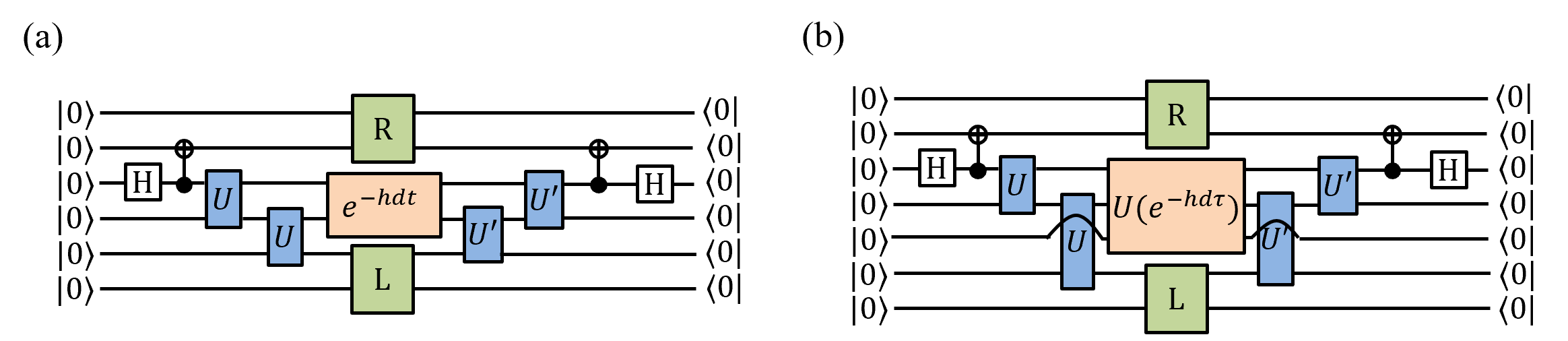}
\caption{(a) Quantum circuit for the cost function in Appendix~\ref{appdx:approch_non_u}, in which the ITE operator is replaced by $e^{-iAd\tau}$. (b) Corresponding circuit using the unitary embedding described in Appendix~\ref{appdx:expand_non_u}.
}
\label{appdx_fig:cost_circ}

\end{figure*}

To construct the environment gates $L$ and $R$ in Fig.~\ref{appdx_fig:cost_circ}, we first convert the unitary gate $U$ into a uMPS unit cell and solve the fixed-point equations shown in Fig.~\ref{appdx_fig:unitary_env}(a) to obtain the environment tensors $l$ and $r$. In practice, these tensors are the leading left and right eigenvectors of the transfer matrix. Because the evolution uses a finite time step, the leading eigenvalue is slightly smaller than 1 rather than exactly equal to 1. Once the tensor $l_{\alpha,\beta}$ [Fig.~\ref{fig:l_L}(a)] has been obtained, the gate $L_{\alpha,\gamma,\beta,\delta}$ [Fig.~\ref{fig:l_L}(b)] must satisfy
\begin{equation}
    \begin{cases}
      \text{$L$ is unitary, namely} & L L^\dagger=I\\
      L_{\alpha,\gamma=0,\beta,\delta=0}=l_{\alpha,\beta}
    \end{cases}    
    \label{appdx_eq:L_cond}.
\end{equation}
Applying the construction above yields the unitary environment gate $L$; the right environment gate $R$ is obtained analogously. The resulting gates satisfy the fixed-point equations shown in Fig.~\ref{appdx_fig:unitary_env}(b) and (c), so the non-unitary environment tensors $l$ and $r$ can be implemented in a quantum circuit at the cost of one ancilla qubit.

The same approach embeds the imaginary-time evolution operator in a unitary gate. Because this construction requires all singular values to be at most 1, we rescale the ITE operator by a sufficiently large positive constant $\alpha$, consistent with the rescaling $M\rightarrow M/\alpha$ in Eq.~\eqref{eq:UofM}:
\begin{equation}
e^{-d\tau H}=WSV^\dagger \rightarrow \frac{e^{-d\tau H}}{\alpha}=W\!\left(\frac{S}{\alpha}\right)\!V^\dagger.
\end{equation}
The overall factor can be removed when the evolved state is normalized. After embedding the ITE operator, the circuit in Fig.~\ref{appdx_fig:cost_circ}(a) is transformed into that in Fig.~\ref{appdx_fig:cost_circ}(b).

\section{Approximation of a non-unitary operator by a unitary operator}
\label{appdx:approch_non_u}
\begin{figure}[ht]
    \includegraphics[width=\columnwidth]{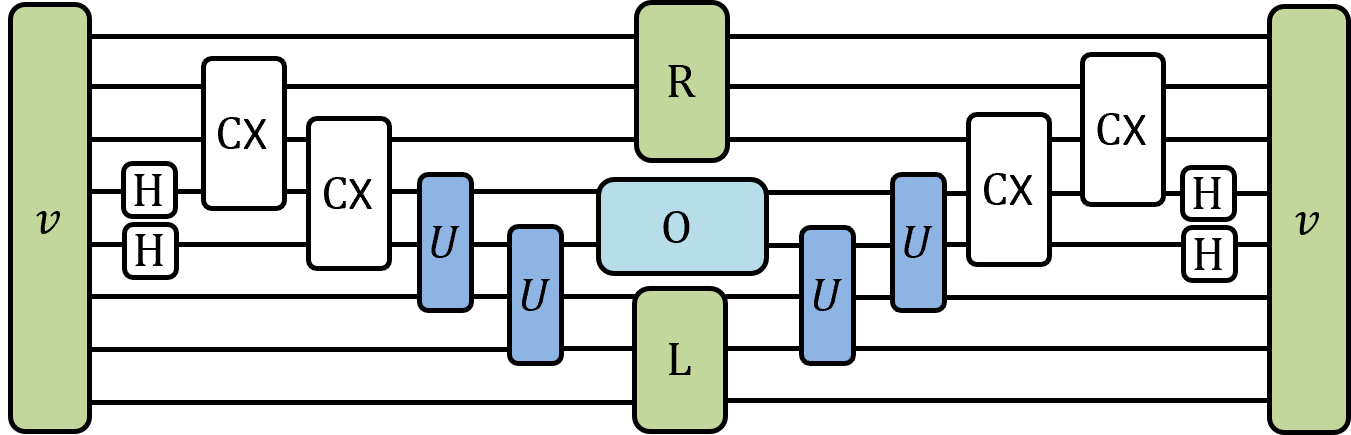}
	\centering
	\caption{TN representation of $\langle\psi(\tau)|O|\psi(\tau)\rangle$, used to optimize $A(\tau)$ in Eq.~\eqref{appdx_eq:A_eq}. Here, $CX$ denotes the CNOT gate, represented as a tensor. The input and output tensors $v$ represent the all-zero state $|000\cdots\rangle$: every element is zero except the element whose indices are all zero, which equals one.}
	\label{appdx_fig:A_cost}
\end{figure}

As discussed in Sec.~\ref{sec:non_unitary}, we approximate the non-unitary operator with a unitary one:
\begin{equation} 
e^{-iA(\tau)d\tau}|\psi(\tau)\rangle \approx e^{-Hd\tau}|\psi(\tau)\rangle,
\end{equation}
where $A(\tau)$ is Hermitian. We determine $A(\tau)$ by minimizing
\begin{widetext}
\begin{equation} \label{appdx_eq:A_eq}
\begin{split} 
A(\tau) 
&= \mathop{\arg\min}_{A(\tau)}{\left\|\frac{e^{-Hd\tau}|\psi(\tau)\rangle}{\sqrt{\langle\psi(\tau)|e^{-2Hd\tau}|\psi(\tau)\rangle}}-e^{-iA(\tau)d\tau}|\psi(\tau)\rangle\right\|^{2}} \\
&= \mathop{\arg\min}_{A(\tau)}{\left[\,2-\frac{1}{\sqrt{\mathcal{N}}}\langle\psi(\tau)|e^{-Hd\tau}e^{-iA(\tau)d\tau}+e^{iA(\tau)d\tau}e^{-Hd\tau}|\psi(\tau)\rangle\right]\,} \\
&= \mathop{\arg\max}_{A(\tau)}{\left[\,\langle\psi(\tau)|e^{-Hd\tau}e^{-iA(\tau)d\tau}+e^{iA(\tau)d\tau}e^{-Hd\tau}|\psi(\tau)\rangle\right]\,}\\
&= \mathop{\arg\max}_{A(\tau)}{\left[\,\langle\psi(\tau)|O|\psi(\tau)\rangle\right]\,},
\end{split}
\end{equation}
\end{widetext}
where $\mathcal{N}=\langle\psi(\tau)|e^{-2Hd\tau}|\psi(\tau)\rangle$ is independent of $A(\tau)$ and $O=e^{-Hd\tau}e^{-iA(\tau)d\tau}+e^{iA(\tau)d\tau}e^{-Hd\tau}$; minimizing the squared distance is equivalent to minimizing the distance itself, and, since $\sqrt{\mathcal{N}}>0$ does not depend on $A(\tau)$, to maximizing $\langle\psi(\tau)|O|\psi(\tau)\rangle$. Because $e^{-Hd\tau}$ is non-unitary and does not preserve the state norm, normalization is required in the first line of Eq.~\eqref{appdx_eq:A_eq}. The operator $O$ is also non-unitary, so $\langle\psi(\tau)|O|\psi(\tau)\rangle$ cannot be evaluated directly as a quantum circuit. Instead, we evaluate it using the TN shown in Fig.~\ref{appdx_fig:A_cost}. 

We parametrize $A$ as a $d^2\times d^2$ Hermitian matrix, where $d$ is the physical dimension. For $d=2$, the matrix is
\begin{equation}
\begin{pmatrix}
p_0      & p_1+ip_2 & p_3+ip_4   & p_5+ip_6 \\
p_1-ip_2 & p_7      & p_8+ip_9   & p_{10}+ip_{11} \\
p_3-ip_4 & p_8-ip_9 & p_{12}     & p_{13}+ip_{14} \\
p_5-ip_6 & p_{10}-ip_{11} & p_{13}-ip_{14} & p_{15}

\end{pmatrix}.
\end{equation}
We use the \texttt{SciPy} optimization routine to find the parameters $[\,p_0,\ldots,p_{15}\,]$ that maximize $\langle\psi(\tau)|O|\psi(\tau)\rangle$, evaluated using the TN in Fig.~\ref{appdx_fig:A_cost}. After obtaining the optimal $A$ from Eq.~\eqref{appdx_eq:A_eq}, we construct the circuit shown in Fig.~\ref{appdx_fig:cost_circ}(a) and use it to find the optimal $U'$ for the next time step.

\begin{figure*}
\centering
  \includegraphics[width=\linewidth]{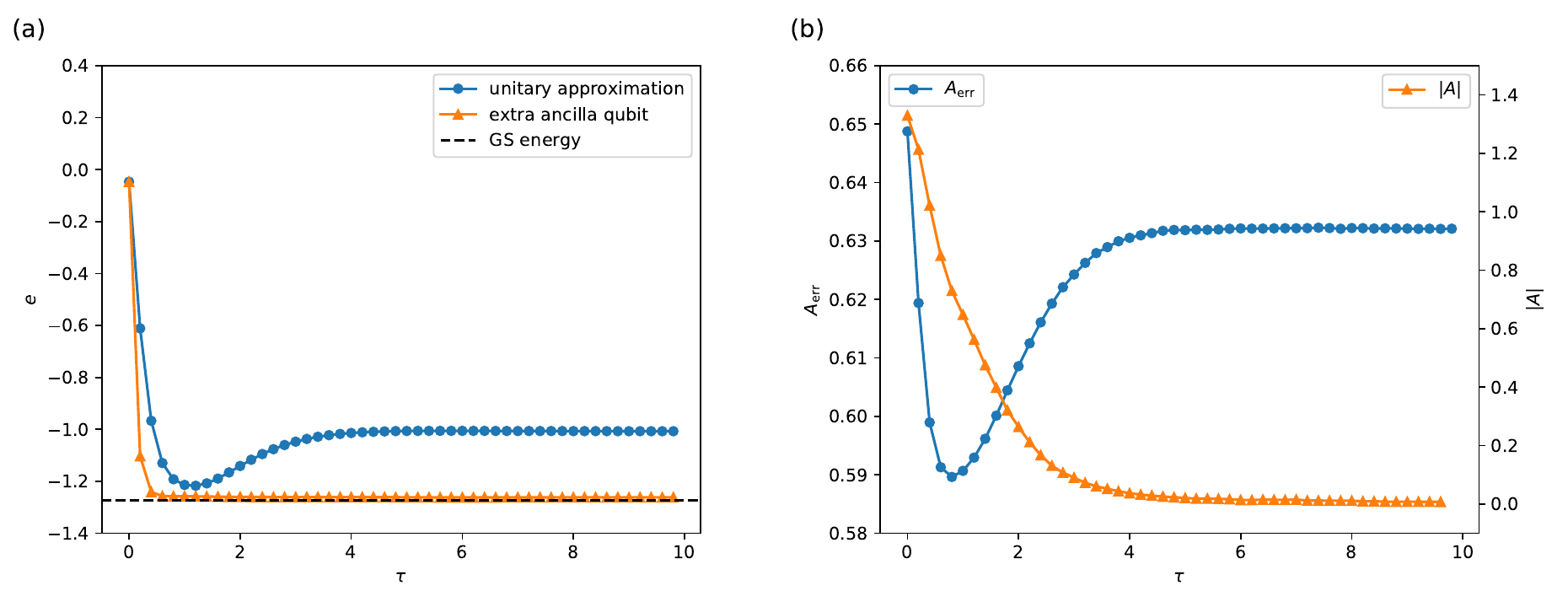}
\caption{(a) Comparison of the methods introduced in Appendices~\ref{appdx:approch_non_u} and \ref{appdx:expand_non_u}. (b) Approximation error defined by the cost function in Fig.~\ref{appdx_fig:A_cost}, measuring the difference between the normalized ITE state $e^{-Hd\tau}|\psi(\tau)\rangle$ and the unitary approximation $e^{-iA(\tau)d\tau}|\psi(\tau)\rangle$.}
\label{appdx_fig:A_err}
\end{figure*}

Figure~\ref{appdx_fig:A_err}(a) compares the statevector results with those obtained using the ancilla-assisted embedding of Appendix~\ref{appdx:expand_non_u}. As discussed in Sec.~\ref{sec:res}, approximating the ITE operator by a unitary operator produces an unexpected increase in energy. To investigate this behavior, we also compute the approximation error
\begin{equation} 
\Delta(\tau) 
= \|\frac{e^{-Hd\tau}|\psi(\tau)\rangle}{\sqrt{\langle\psi(\tau)|e^{-2Hd\tau}|\psi(\tau)\rangle}}-e^{-iA(\tau)d\tau}|\psi(\tau)\rangle\|,
\end{equation}
shown in Fig.~\ref{appdx_fig:A_err}(b). The approximation error and energy exhibit similar trends, suggesting that the energy increase is caused by the unitary approximation.

\section{Statistics of the energy distribution from measurement-based updates}
\label{appdx:energy_stat}

We analyze the energy distribution in Fig.~\ref{fig:e_evo_statistics}(b) at the level of individual trajectories. We first formulate the simplest mechanism that could produce the Maxwell--Boltzmann-like form in Eq.~\eqref{eq:maxwell_boltzmann}: Gaussian parameter fluctuations about a single optimum in a locally quadratic landscape. We then derive testable predictions and compare them with the ensemble of $100$ trajectories. The data do not support the single-optimum picture; instead, the distribution is the envelope of an inhomogeneous ensemble of stalled optimization points.

\subsection{A single-optimum \texorpdfstring{$\chi^2$}{chi-squared} hypothesis}
\label{appdx_sub:chi2_hypothesis}

Let $\boldsymbol{\theta}\in\mathbb{R}^{d}$ denote the PQC parameters and $\boldsymbol{\theta}_\star$ a local minimizer of the energy. For the ansatz used here, $d=15$.
Near $\boldsymbol{\theta}_\star$, the energy may be approximated by a quadratic form
\begin{equation}
E(\boldsymbol{\theta}) \;\approx\; E_0 + \tfrac12\,(\boldsymbol{\theta}-\boldsymbol{\theta}_\star)^{\!\top} 
\mathbf{H}\,(\boldsymbol{\theta}-\boldsymbol{\theta}_\star), \qquad \mathbf{H}\succ 0,
\label{eq:quad}
\end{equation}
where $\mathbf{H}$ is the local Hessian. 
If the measurement-based update is modeled as a noisy, mean-reverting (discrete-time Ornstein--Uhlenbeck) process,
\begin{equation}
\boldsymbol{\theta}_{t+1}-\boldsymbol{\theta}_\star \;=\; (\mathbf{I}-\eta\,\mathbf{H})\,(\boldsymbol{\theta}_{t}-\boldsymbol{\theta}_\star)
\;+\; \boldsymbol{\xi}_t,
\qquad \boldsymbol{\xi}_t \sim \mathcal{N}(\mathbf{0},\, 2\mathcal{D}\,\mathbf{I}),
\label{eq:ou}
\end{equation}
with effective step size $\eta>0$ and noise strength $\mathcal{D}>0$ determined by finite-sampling noise, then at stationarity $\boldsymbol{\theta}-\boldsymbol{\theta}_\star$ is Gaussian with covariance $\boldsymbol{\Sigma}$. For small $\eta$, the covariance obeys the balance condition
\begin{equation}
\eta\,\mathbf{H}\,\boldsymbol{\Sigma} \;+\; \boldsymbol{\Sigma}\,\eta\,\mathbf{H} \;=\; 2\mathcal{D}\,\mathbf{I}
\;\;\Rightarrow\;\; 
\boldsymbol{\Sigma} \;=\; a\,\mathbf{H}^{-1}, 
\qquad a \equiv \mathcal{D}/\eta.
\label{eq:sigma}
\end{equation}
In the whitened coordinates $\mathbf{q}\equiv\mathbf{H}^{1/2}(\boldsymbol{\theta}-\boldsymbol{\theta}_\star)$, one has $\mathbf{q}\sim\mathcal{N}(\mathbf{0},a\mathbf{I})$, and therefore
\begin{equation}
\Delta E \;\equiv\; E(\boldsymbol{\theta})-E_0 \;=\; \tfrac12\,\|\mathbf{q}\|_2^2,
\label{eq:gamma}
\end{equation}
which follows a Gamma distribution with shape $d_\mathrm{eff}/2$ and scale $a$, where $d_\mathrm{eff}\le d$ counts the effectively fluctuating quadratic modes.\footnote{If the fluctuations are anisotropic after whitening, $\Delta E$ follows a generalized $\chi^2$ (a sum of unequally weighted $\chi^2$ variables) rather than a single Gamma distribution.}
For $d_\mathrm{eff}=3$, this Gamma distribution reduces exactly to the Maxwell--Boltzmann-like form of Eq.~\eqref{eq:maxwell_boltzmann} with $x=E-E_0$.

This hypothesis makes two sharp predictions. 
First, the floor $E_0$ of the distribution should coincide with the variational minimum. Second, independently of $E_0$, the \emph{within-trajectory} energy fluctuations in the stationary regime should themselves follow a $\chi^2$ distribution. For $d_\mathrm{eff}=3$, the standardized fluctuations should have skewness $\sqrt{8/d_\mathrm{eff}}\approx1.63$ and excess kurtosis $12/d_\mathrm{eff}=4$.

\subsection{Trajectory-resolved analysis}
\label{appdx_sub:trajectory_analysis}

Both predictions fail when tested against the $100$ trajectories underlying Fig.~\ref{fig:e_evo_statistics}, using the last $10$ time steps of each run ($1000$ points in total).

\emph{Stall offset.} 
The lowest energy attained across all $1000$ points is $e_{\min} \simeq -1.260$, which lies approximately $0.013$ \emph{above} the $D=2$ variational minimum $e_0^{(D=2)} \simeq -1.2725$. 
Every trajectory therefore stalls before reaching the optimum, and the floor of the observed distribution is a stall threshold rather than the ground-state energy, contradicting the first prediction. 
The magnitude of the offset is consistent with the termination condition of the optimizer: the shot-noise scale of a single cost estimate, $\sigma \simeq 3\times 10^{-3}$ from Eq.~\eqref{eq:std} at $P_0 \simeq 0.07$ and $N_{\mathrm{shots}}=8192$, is comparable to the \texttt{COBYLA} tolerance of $\sim\!10^{-3}$, so the iteration terminates once cost differences are buried in sampling noise, before the local gradient vanishes.

\emph{Ensemble inhomogeneity.} 
Classifying a run as converged if the mean of its last $10$ energies lies below $e=-1.05$, we find that $16$ of the $100$ runs fail to converge; these runs account for the secondary structure near $e\approx -0.94$ and for the entire high-energy tail of the distribution. 
Among the $84$ converged runs, the run-to-run standard deviation of the plateau energies ($\approx 0.046$) is comparable to the typical fluctuation within a single run ($\approx 0.05$), so that roughly $40\%$ of the variance in the late-time window is between-run variance. 
The $1000$ points are, moreover, strongly correlated within runs (lag-one autocorrelation $\approx 0.5$), so that confidence intervals treating them as independent are substantially overconfident.

\emph{Within-run statistics.} 
Pooling the standardized within-run fluctuations of the converged runs yields a skewness of approximately $0.5$ and an excess kurtosis of approximately $-0.7$, in clear disagreement with the $\chi^2_3$ predictions of $1.63$ and $4$. The fluctuations within an individual trajectory are therefore approximately Gaussian, contradicting the second prediction.

\begin{figure*}
\centering
  \includegraphics[width=\linewidth]{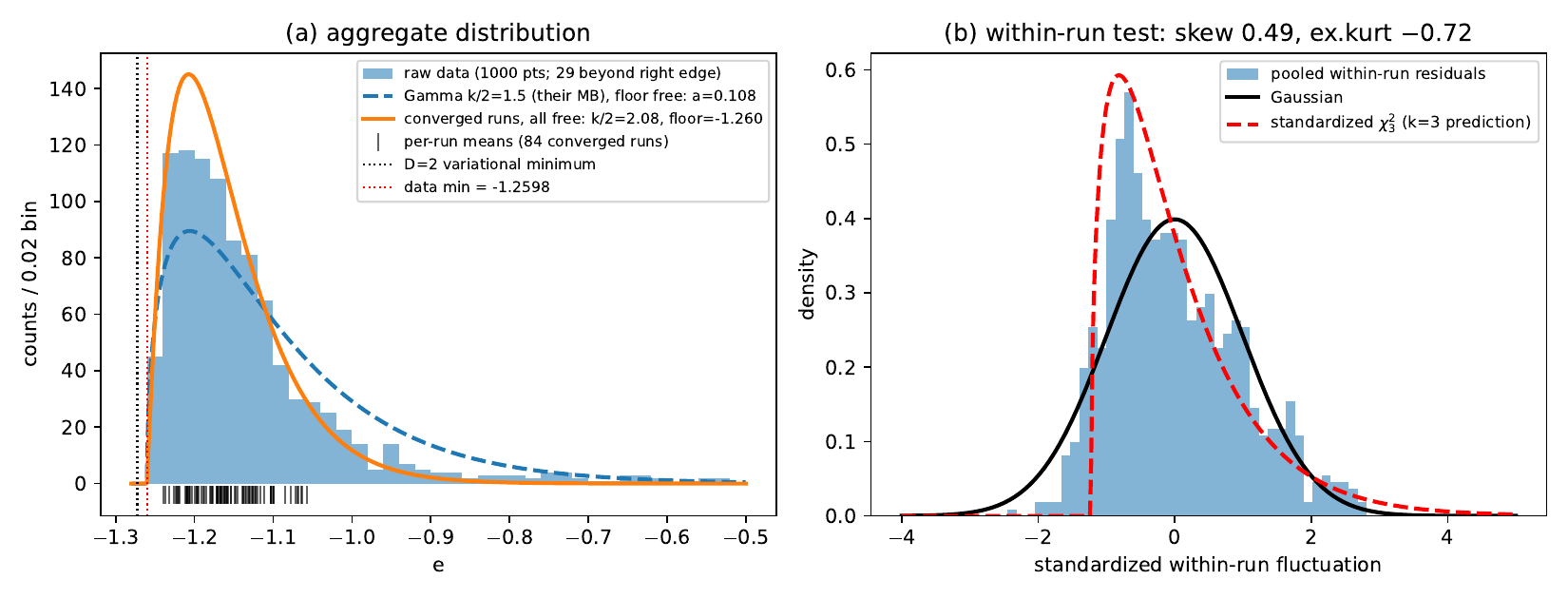}
\caption{(a) Energy distribution from Fig.~\ref{fig:e_evo_statistics}(b), compared with the $k=3/2$, $a=0.08$, $E_0=-1.28$ curve of Eq.~\eqref{eq:maxwell_boltzmann} and a Gamma fit to the converged runs with free shape, scale, and floor. The rug marks show the mean late-time energy of each converged run; vertical lines mark the $D=2$ variational minimum and the lowest attained energy. (b) Pooled standardized within-run fluctuations of the converged runs, compared with a Gaussian distribution and the standardized $\chi^2_3$ distribution predicted by the single-optimum hypothesis. The pooled fluctuations have skewness $0.49$ and excess kurtosis $-0.72$.}
\label{appdx_fig:e_stat}
\end{figure*}

Consistent with this model mismatch, a maximum-likelihood fit of the generalized form $f(x)\propto x^{k-1}e^{-x/a}$, a Gamma distribution with free shape parameter $k$, does not yield a robust value. Depending on whether the floor is fitted freely or fixed at $e_0^{(D=2)}$ and whether the unconverged runs are included, the fitted shape varies from $k\approx1$ to $k\approx3$. For the converged subset alone, a run-level bootstrap gives a $95\%$ interval of approximately $[1.5,2.3]$; see Fig.~\ref{appdx_fig:e_stat}(a). Indeed, the fitted floor $E_0=-1.28$ of the $k=3/2$ curve of Eq.~\eqref{eq:maxwell_boltzmann} in Fig.~\ref{appdx_fig:e_stat}(a) lies below the variational minimum---impossible for variational energies---confirming that the $k=3/2$ form describes only the envelope. 
The value $k=3/2$ in Eq.~\eqref{eq:maxwell_boltzmann} is therefore an adequate empirical description of the envelope but does not identify an underlying $\chi^2$ law with three degrees of freedom.

\subsection{Interpretation: an ensemble of stalled fixed points}
\label{appdx_sub:stall}

The three observations above, summarized in Fig.~\ref{appdx_fig:e_stat}, admit a consistent interpretation. 
Because each trajectory terminates at a point where the optimizer tolerance meets the sampling noise, the stall point is \emph{not} a stationary point of the energy landscape, and the local gradient $\mathbf{g}=\nabla E$ does not vanish there. 
The leading energy fluctuation within a run is then \emph{linear} in the Gaussian parameter noise, $\delta E\approx\mathbf{g}\cdot\delta\boldsymbol{\theta}$, and is therefore Gaussian, as observed, rather than following the quadratic form in Eq.~\eqref{eq:gamma}. 
The right-skewed, one-sided shape of the aggregate distribution instead originates from the run-to-run distribution of stall energies above the variational minimum, convolved with the approximately Gaussian within-run fluctuations and augmented by the unconverged subpopulation. 
This picture makes a falsifiable prediction that separates the two contributions: the within-run fluctuation width should scale as $1/\sqrt{N_{\mathrm{shots}}}$, whereas the distribution of stall offsets should be controlled primarily by the optimizer tolerance and the time step $d\tau$; verifying these scalings is left for future work.

\section{Derivation of the \texorpdfstring{$\sqrt{2}$}{square-root-of-two} ratio}
\label{appdx:sqrt2}

Let $P(\boldsymbol{\theta})$ denote the cost function, such as $P_0$, and consider fluctuations along the dominant one-dimensional mode near the operating point in the stationary regime. Denoting this mode by $\theta\in\mathbb{R}$, we linearize
\begin{equation}
P(\theta) \;\approx\; P_\star + s\,\delta\theta, \qquad \delta\theta := \theta-\theta_\star,\quad s:=\left.\frac{dP}{d\theta}\right|_{\theta_\star}.
\label{eq:linP}
\end{equation}
We note that a nonvanishing local sensitivity $s$ is consistent with the identification of the operating point as a non-stationary stall point in Sec.~\ref{appdx_sub:stall}.

\emph{Fixed-circuit case.} Estimating $P$ from $N_{\mathrm{shots}}$ samples while holding the circuit fixed produces binomial shot noise:
\begin{equation}
\hat{P}_{\mathrm{sv}} = P(\theta_\star) + \zeta,\;  \mathbb{E}[\zeta]=0,\;
\mathrm{Var}(\zeta) = \sigma_{\mathrm{sv}}^2 := \frac{P_\star(1-P_\star)}{N_{\mathrm{shots}}}.
\label{eq:svvar}
\end{equation}

\emph{Measurement-updated case.} The parameters are iteratively updated using noisy finite-shot estimates, after which the cost is measured again:
\begin{align}
\delta\theta_{t+1} &= (1-\eta\lambda)\,\delta\theta_t \;-\; \eta\,\varepsilon_t, 
\qquad \varepsilon_t \sim \mathcal{N}(0,\sigma_g^2), 
\label{eq:ou_scalar}\\
\hat{P}_{t+1} &= P(\theta_{t+1}) + \zeta_{t+1} \;\approx\; P_\star + s\,\delta\theta_{t+1} + \zeta_{t+1},
\end{align}
where $\lambda>0$ is the local mean-reversion rate, $\eta>0$ is the step size, and $\sigma_g^2$ characterizes the update noise induced by finite-shot sampling.\footnote{For update rules based on finite differences of estimates obtained with $N_{\mathrm{shots}}$ samples, one typically has $\sigma_g^2=\gamma\sigma_{\mathrm{sv}}^2$ with $\gamma=O(1)$; $\gamma$ absorbs details of the probing scheme.}

At stationarity, Eq.~\eqref{eq:ou_scalar} defines a first-order autoregressive process with variance
\begin{equation}
\mathrm{Var}(\delta\theta) \;=\; \frac{\eta^2 \sigma_g^2}{1-(1-\eta\lambda)^2}
\;\approx\; \frac{\eta\,\sigma_g^2}{2\lambda}\quad(\text{small }\eta).
\label{eq:vardth}
\end{equation}
Thus the variance of the measured cost in the updated run is
\begin{eqnarray}
    \mathrm{Var}(\hat{P}_{\mathrm{upd}}) &=& \underbrace{\sigma_{\mathrm{sv}}^2}_{\text{current shot noise}}
\;+\;
\underbrace{s^2\,\mathrm{Var}(\delta\theta)}_{\text{propagated param.~noise}}\nonumber\\
&\approx&
\sigma_{\mathrm{sv}}^2 \;+\; s^2\,\frac{\eta\,\sigma_g^2}{2\lambda}.
\label{eq:var_update}
\end{eqnarray}
Writing $\sigma_g^2 = \gamma\,\sigma_{\mathrm{sv}}^2$ with $\gamma=O(1)$, the ratio of standard deviations is
\begin{equation}
\mathcal{R} \;:=\; \frac{\sigma_{\mathrm{upd}}}{\sigma_{\mathrm{sv}}}
\;=\;
\sqrt{1 + \frac{s^2\,\eta\,\gamma}{2\lambda}}
\;\;\; \xrightarrow[\;\;s^2\eta\gamma/(2\lambda)\;\to\;1\;\;]{}\;\;\; \sqrt{2}.
\label{eq:ratio}
\end{equation}
Equation~\eqref{eq:ratio} shows that the updated run contains two independent sources of variance: the current measurement shot noise $\sigma_{\mathrm{sv}}^2$ and the propagated noise $s^2\mathrm{Var}(\delta\theta)$ from parameter updates driven by previous noisy measurements.
When the stationary parameter fluctuations are such that $s^2\,\mathrm{Var}(\delta\theta)\approx\sigma_{\mathrm{sv}}^2$,
the two contributions add in quadrature, giving
$\sigma_{\mathrm{upd}}^2 \approx 2\sigma_{\mathrm{sv}}^2$ and hence $\sigma_{\mathrm{upd}}/\sigma_{\mathrm{sv}}\approx \sqrt{2}$,
which is what we observe in the saturation regime of Fig.~\ref{fig:meas_update}(a). 
The condition $s^2\eta\gamma/(2\lambda)\approx 1$ is not imposed by the derivation; rather, it emerges as an attractor of the stall dynamics. 
By Eq.~\eqref{eq:vardth}, the condition is equivalent to $s^2\,\mathrm{Var}(\delta\theta)\approx\sigma_{\mathrm{sv}}^2$, i.e., the cost fluctuation induced by the stationary parameter scatter equals the shot noise of a single estimate. 
A trust-region method such as \texttt{COBYLA} shrinks its step scale $\rho$ as long as it can resolve improvement, and loses this ability precisely when the cost change produced by a step of size $\rho$ falls to the noise floor, $|s|\rho\sim\sigma_{\mathrm{sv}}$; the stationary parameter scatter is thereby pinned at $\delta\theta\sim\sigma_{\mathrm{sv}}/|s|$, which enforces $s^2\,\mathrm{Var}(\delta\theta)\sim\sigma_{\mathrm{sv}}^2$ automatically. 
This is the same signal-equals-noise condition that produces the stall offset of Sec.~\ref{appdx_sub:stall}: the energy floor and the variance ratio are two manifestations of a single termination mechanism. 
The argument fixes this balance only up to an $O(1)$ factor determined by the acceptance rule and probing scheme. Thus, $\mathcal{R}\approx\sqrt{2}$ should not be interpreted as a universal constant. Because the acceptance test acts on the total cost change, the total induced variance saturates at $\sigma_{\mathrm{sv}}^2$, making the prediction independent of the number of fluctuating modes. The mechanism makes two falsifiable predictions. First, because both variance contributions scale with $\sigma_{\mathrm{sv}}^2$, the ratio should be invariant under changes in $N_{\mathrm{shots}}$. Second, the prediction should fail when the optimizer tolerance greatly exceeds the shot-noise scale, making termination tolerance-limited, or for schemes with decaying step sizes, for which $\mathcal{R}\to1$. 
In the present simulations, the tolerance ($\sim\!10^{-3}$) lies just below the shot-noise scale ($\sigma_{\mathrm{sv}}\simeq 3\times 10^{-3}$), placing the optimization in the noise-limited regime where the argument applies, consistent with the observed $\sqrt{2}$.

\end{document}